\documentclass[aps,prd,twocolumn,showpacs,floatfix,preprintnumbers,amsmath,amssymb,nofootinbib,groupedaddress]{revtex4-1}

\input epsf
\usepackage{graphicx}
\usepackage{color}
\usepackage{mathtools}

\usepackage[utf8]{inputenc}

\def\VEV#1{\left\langle #1 \right\rangle}

\newcommand{\be}{\begin{equation}}
\newcommand{\ee}{\end{equation}}

    \newcommand{\ba}{\begin{align}}
  \newcommand{\ea}{\end{align}} 

\newcommand{\vdm}{v_{\chi}}
\newcommand{\Vdm}{V_{\chi b}}

\newcommand{\mdm}{m_{\chi}}

\begin{document}

\title{Heating of baryons due to scattering with dark matter during the dark ages}

\author{Julian B. Mu\~noz, Ely D. Kovetz, and Yacine Ali-Ha\"imoud}
\affiliation{Department of Physics and Astronomy, Johns
     Hopkins University, 3400 N.\ Charles St., Baltimore, MD 21218}

\date{\today}

\begin{abstract}
We explore the effects of elastic scattering between dark matter and baryons on the 21-cm signal
during the dark ages. In particular, we consider a dark-matter--baryon interaction with a cross section
of the form $\sigma = \sigma_0 v^{-4}$, in which case the effect of the drag force between 
the dark matter and baryon fluids grows with time. We show that, as opposed to what was previously thought, this effect heats up the baryons due to the relative velocity between dark matter and baryons. This creates an additional source of fluctuations, which can potentially make interactions easier to detect by 21-cm measurements than by using the cosmic microwave 
background and the Lyman-$\alpha$ forest. 
Our forecasts show that the magnitude of the cross section can be probed to $\sigma_0\sim 3\times 10^{-42}$ cm$^2$ for $\mdm \ll 1$ GeV and $\sigma_0\sim 2 \times 10^{-41}\ (\mdm/10\, \rm GeV)$ cm$^2$ for $\mdm \gg 1$ GeV with next generation experiments, and improved to $\sigma_0\sim 4\times  10^{-44}$ cm$^2$ for $\mdm \ll 1$ GeV and $\sigma_0\sim 4 \times 10^{-43}\ (\mdm/10\, \rm GeV)$ cm$^2$ for $\mdm\gg  1$ GeV with futuristic experiments.
\end{abstract}

\maketitle

\section{Introduction}

The standard picture of cold dark matter (CDM) \cite{Peebles:1982ff} seems to fit very well with our current observational constraints \cite{1502.01589}. There are, however, a few puzzles that would require dark matter to have non-zero interactions \cite{1103.0007,astro-ph/9810070,astro-ph/0612410,astro-ph/9909386}. Moreover, several models for the dark-matter (DM) particle predict some level of weak non-gravitational interaction with standard-model baryons~\cite{astro-ph/0406355,hep-ph/0003010}. Here we will study these interactions during the dark ages.

The simplest way to observe these interactions would be through direct detection experiments, such as DarkSide \cite{1410.0653,1501.03541}, LUX \cite{1310.8214} and XENON100 \cite{1207.5988}. These experiments are very sensitive to large dark-matter masses but cannot constrain interactions for DM masses below $\sim 10$ GeV due to the small recoil of the nuclei in any given interaction. 
A different probe would consist of indirect early-time effects of these interactions. One example would be the modification of the small-scale power spectrum, due to the drag induced in the DM by the interactions \cite{1311.2937}, which would be observable in the cosmic microwave background (CMB), as well as in Ly-$\alpha$ forest measurements. Another example is CMB spectral distortions, which would be created by the indirect coupling, through baryons, of dark matter and photons in the very early universe \cite{1506.04745}. 
These last two probes require interactions to be relevant at early times, so they are not sensitive to
all velocity dependences. Some models for dark-matter--baryon interactions may elude constraints because interactions get stronger at later times. We will focus on one of those models, in which the interaction cross section is parametrized by $\sigma=\sigma_0 v^{-4}$, one realization of which would be dark-matter milicharge \cite{astro-ph/0406355}. To constrain interactions at later times, a useful probe is the 21-cm line during the dark ages.

The dark ages are the period following primordial recombination and preceding the formation of the first luminous objects. During this cosmic era the only known observable is the redshifted Hydrogen hyperfine transition, which traces the neutral Hydrogen density \cite{astro-ph/0312134}. This observable has been proposed as a probe of  non gaussianities \cite{astro-ph/0610257,1506.04152}, as well as other effects that would modify the small-scale power spectrum \cite{astro-ph/0702600}. Furthermore, it has recently been proposed for the study of DM-baryon interactions \cite{1408.2571}. We go beyond the analysis in Ref.~\cite{1408.2571} by including the effect of relative velocities, which turns out to change the results significantly.

Interactions between baryons and dark matter can be detected through their effect on the brightness temperature of the 21-cm line. This brightness temperature is proportional to the difference between the spin temperature of the neutral Hydrogen and the CMB temperature. In the standard scenario the spin temperature is coupled to the baryon temperature during the redshift range $z\sim 30-200$. This creates a departure between spin temperature and CMB temperature. As shown in Ref. \cite{1408.2571}, if the baryons are cooled down (by interacting with a colder fluid, like the dark matter) the spin temperature will be lower, modifying the overall brightness temperature. 

We emphasize that these interactions do not cause just cooling of the baryons, but also heating.
In the usual picture of interaction between two fluids, the warmer fluid will lose energy toward heating up the colder one, while there will be no energy transfer if both fluids have the same temperature. However, if there is a relative velocity between the two fluids -dark matter and baryons in our case- there will be an additional friction term that will tend to damp this relative velocity. The kinetic energy lost in this manner will induce heating in both fluids. The magnitude of this effect depends on the initial relative velocity, which is given by a Gaussian variable with a (3D) variance of $\sim 29$ km/s at kinematic decoupling ($z\approx1010$) \cite{1005.2416,1312.4948}.

The brightness temperature will then acquire an additional spatial dependence, through the local variation of the relative velocities.
Quantifying this effect, we find that during the dark ages it 
creates an additional contribution to the power spectrum  of 21-cm temperature fluctuations, which can be more than an order of magnitude bigger at large scales than the standard one, even for values of the cross section allowed by current CMB studies \cite{1311.2937}. We study the detectability of this new signal with an SKA-inspired interferometer\footnote{https://www.skatelescope.org/.} and with a more futuristic proposed experiment. We also study how the global signal changes due to interactions and discuss the prospects for experiments such as NenuFAR\footnote{http://nenufar.obs-nancay.fr.}.

This paper is organized as follows. In Section \ref{sec:evol} we derive the drag and heating terms and find their effects on the dark-matter and baryon temperatures. Later, in Section~\ref{sec:21cm} we study how the change in the baryon temperature affects the signal of the 21-cm line during the dark ages. In Section~\ref{sec:detection}  we carry out a detectability analysis of this signal. We discuss some generalizations, as well as other possible effects of the interactions, in Section~\ref{sec:discussion} before drawing our conclusions in Section~\ref{sec:conc}.

\section{Evolution of interacting dark matter and baryon fluids}
\label{sec:evol}

In this section we will study how the interactions between DM and baryons change their temperatures. To do that we will have to calculate the drag on the relative velocity due to interactions with baryons, as well as the heating effect on both fluids. Our results will rely on the current understanding of relative velocities, so let us start with a brief review.

\subsection{Velocities}

In the standard cosmological evolution, dark matter starts collapsing as soon as matter-radiation equality is reached. Baryons, however, cannot cluster due to radiation pressure, until they decouple from the photon background. This difference in their evolution history generates a relative velocity between the two components.  After the baryons and photons kinematically decouple, at redshift $z\approx 1010$, this velocity redshifts away, since the baryons experience infall into the DM gravitational wells. Ref.~\cite{1005.2416} first pointed out that relative velocities affect the formation of small-scale structure. Their effect on the standard power spectrum of 21-cm fluctuations in the dark ages was studied in Ref.~\cite{1312.4948}. 

At kinematic decoupling, the relative velocities $\mathbf V_{\chi b} \equiv \mathbf V_{\chi} -\mathbf V_{b}$ follow a Gaussian distribution, where $\mathbf V_{\chi}$ and $\mathbf V_{b}$ are the DM and baryon bulk velocities. Then the differential probability of having an initial relative velocity $\mathbf V_{\chi b,0}$ is given by
\be
\mathcal P(\mathbf V_{\chi b,0}) = \dfrac{e^{-3 \mathbf V_{\chi b,0}^2/(2V_{\rm rms}^2)}}{(\frac{2\pi}{3} V_{\rm rms}^2)^{3/2} },
\label{eq:pv}
\ee
where the value of the (3D) width of this distribution is $V_{\rm rms}=29$ km/s $\sim 10^{-4}\,c$ at kinematic decoupling ($z=1010$) \cite{1312.4948}. 
This rms value as well as the full power spectrum of $V_{\chi b, 0}$ can be simply extracted from standard linear Boltzmann codes \cite{astro-ph/0702600,1104.2933}.

Elastic interactions between fluids with a relative velocity will have two different effects. First, they will tend to decrease the relative velocity and achieve mechanical equilibrium, which in our scenario will manifest itself as a drag on the relative velocity \cite{1311.2937}. Second, they will thermally couple the fluids, tending to equilibrate their temperatures.

We start by calculating the drag on the relative velocity.

\subsection{Drag term}

Throughout the text we consider cross sections parametrized as $\sigma=\sigma_0 v^{-4}$. First we analyze the velocity change due to the collision with a baryon with velocity $\mathbf v_b$. In the center-of-mass (CM) frame the initial velocity of the DM particle will be
\be
\mathbf \vdm^{(\mathrm{CM}),0} = \left( \mathbf\vdm-\mathbf v_b \right) \dfrac {m_b}{m_b+\mdm},
\ee
and in an elastic collision the final velocity can be parametrized by the angle toward which it is scattered, so the final velocity of the dark matter particle is
\be
\mathbf \vdm^{(\mathrm{CM}),f} = \vdm^{(\mathrm{CM}),0} \hat n,
\ee
where $\hat n$ is a unit vector. The change in velocity in a single collision (which is Galilean invariant, and hence frame independent) is
\be
\Delta \mathbf \vdm = \dfrac{m_b}{m_b+\mdm} \left |\mathbf \vdm - \mathbf{v}_b \right | \left ( \hat n - \dfrac{\mathbf \vdm - \mathbf{v}_b}{\left| \mathbf \vdm - \mathbf{v}_b\right|} \right).
\label{eq:deltavchi}
\ee

To calculate the full effect of the interactions we need to include the rate at which interactions happen, and average over the velocities of the fluid elements. 
The rate of interactions in a particular direction $d\hat n$ is $d\sigma/d\hat n \left| \mathbf \vdm - \mathbf{v}_b\right| n_b$,  where $\sigma(\left| \mathbf \vdm - \mathbf{v}_b\right|)$ is the cross section as a function of the relative velocity, and $n_b$ is the number density of baryons (targets). 
The time derivative of the DM bulk velocity will then be
\small
\be
\dfrac{d\mathbf V_\chi}{dt}=n_b \int d^3 v_\chi f_\chi \int d^3 v_b f_b\left| \mathbf \vdm - \mathbf{v}_b\right| \int d \hat n  \dfrac{d\sigma}{d\hat n}  \Delta \mathbf v_\chi,
\label{eq:dvchidt}
\ee
\normalsize
and we can perform the inner integral, by plugging Eq.~(\ref{eq:deltavchi}) into Eq.~(\ref{eq:dvchidt})  and realizing it has to be proportional to the only direction ($ \mathbf \vdm - \mathbf{v}_b$) inside the integral, to find
\small
\be
\dfrac{d\mathbf  V_\chi}{dt}=-\dfrac{\rho_b}{m_b+\mdm} \int d^3 v_\chi f_\chi \int d^3 v_b f_b (\mathbf \vdm - \mathbf{v}_b)\left| \mathbf \vdm - \mathbf{v}_b\right| \bar \sigma,
\label{eq:dvchi}
\ee
\normalsize
where we have defined the momentum-transfer cross section as 
\be
\bar\sigma(\left| \mathbf \vdm - \mathbf{v}_b\right|) \equiv \int d(\cos\theta)  \dfrac{d\sigma}{d \cos\theta} \left ( 1- \cos\theta  \right).
\label{eq:sigmabar}
\ee

Alternatively, we could have calculated the drag on the baryon velocity, which is given by exchanging $\chi \leftrightarrow b$ in Eq.~(\ref{eq:dvchi}), so that $d\mathbf V_b/dt = - (\rho_\chi/\rho_b) d\mathbf V_\chi/dt $. The relative velocity between the two fluids will then evolve as
\small
\be
\dfrac{d\mathbf V_{\chi b}}{dt}=-\dfrac{\rho_m}{m_b+\mdm} \int d^3 v_\chi f_\chi \int d^3 v_b f_b (\mathbf \vdm - \mathbf{v}_b)\left| \mathbf \vdm - \mathbf{v}_b\right| \bar \sigma,
\label{eq:dvchib}
\ee
\normalsize
where we have defined $\rho_m \equiv \rho_b + \rho_\chi$.

To calculate the two integrals over velocities we define two new variables $\mathbf v_m$ and $\mathbf v_{\rm th}$, as
\ba
\mathbf v_m  &\equiv \dfrac{\dfrac{\mdm}{T_\chi} \mathbf v_\chi+\dfrac{m_b}{T_b} \mathbf v_b}{\dfrac{\mdm}{T_\chi}+\dfrac{m_b}{T_b}}, \quad {\rm and} \\ 
\mathbf v_{\rm th}  &\equiv \mathbf v_\chi - \mathbf v_b,
\label{eq:newv}
\end{align}
so that the velocity distributions $f$ factorize
\be
\int d^3 v_\chi f_\chi \int d^3 v_b f_b=\int d^3 v_{\rm th} f_{\rm th} \int d^3 v_m f_m.
\ee
Nothing will depend on $\mathbf v_m$, so we can just integrate it out, leaving then only the integral of the relative velocity $\mathbf v_{\rm th}$. The distribution function $f_{\rm th}$ of this velocity is a Gaussian displaced from the origin by $\mathbf V_{\chi b}$ and with thermal width given by the sums of the baryon and DM widths, $ T_\chi/\mdm + T_b/m_b$. The integral to calculate hence reduces to
\be
\dfrac{d\mathbf V_{\chi b}}{dt}=-\dfrac{\rho_m}{m_b+\mdm} \int d^3 v_{\rm th} f_{\rm th} \mathbf v_{\rm th} v_{\rm th} \bar \sigma(v_{\rm th}),
\label{eq:dvchib2}
\ee

Focusing on the case in which the interaction cross section is parametrized as $\bar \sigma = \sigma_0 v^{-4}$, the drag term is given by
\ba
D(V_{\chi b}) &\equiv -\dfrac{d  V_{\chi b}}{dt}  = \dfrac{\rho_m\sigma_0}{m_b+\mdm} \dfrac 1 { V_{\chi b}^2} F(r),
\label{eq:drag}
\end{align}
where we have defined $r\equiv V_{\chi b}/u_{\rm th}$, and $u_{\rm th}^2\equiv T_b/m_b + T_\chi/m_\chi$, which is the variance of the thermal relative motion of the two fluids. The function $F(r)$ is determined as
\be
F(r)\equiv  \text{erf}\left(\frac{r}{\sqrt{2}}\right) - \sqrt{\dfrac{2}{\pi}} e^{-r^2/2} r ,
\ee
which grows with $r$ from zero at $r=0$ to one at $r\to\infty$.

\subsection{Heating}

We now study the second effect that interactions have on the dark-matter and baryon fluids, namely heating. 
Interactions between two fluids (1 and 2) with different temperatures will tend to heat up the colder fluid (in our case the cold dark matter) at the expense of the energy of the warmer fluid, tending to equalize their temperatures. The heating rate is usually proportional to the temperature difference $(T_1-T_2)$. We will show here that, if there is a relative velocity between the two fluids, the heating rate will also include a friction term that will heat up both fluids, independently of their temperature difference.

There is an intuitive reason to expect a heating term even for equal-temperature fluids, if two fluids with the same temperature collide with a relative velocity, and then equilibrate, this final relative velocity should vanish. The kinetic energy would hence get transformed into a higher final temperature for both fluids, due to conservation of energy.

Let us calculate the heating rate $\dot Q_b$ of the baryons in their instantaneous rest frame, where the change in energy will directly give us the heat instead of having to add bulk motions. A baryon changes its energy in a collision by $\Delta E_b = m_b \mathbf v_{\rm CM} \cdot \Delta\mathbf v_b = -m_\chi \mathbf v_{\rm CM} \cdot \Delta\mathbf v_\chi$ \cite{1311.2937}, where $\mathbf v_{\rm CM} = (m_b \mathbf v_b + m_\chi \mathbf v_\chi)/(m_b + m_\chi)$. The heating of the baryonic fluid per unit time is  
\ba
\dfrac{dQ_b}{dt} =& \dfrac{m_b \rho_\chi}{(\mdm+m_b)} \int d^3 v_b f_b \int d^3 \vdm f_\chi (\vdm)
   \nonumber\\
 & \times \bar \sigma\left(|\mathbf \vdm - \mathbf v_b|\right)  |\mathbf \vdm - \mathbf v_b| \left[ \mathbf v_{\rm CM}\cdot  (\mathbf v_b-\mathbf v_\chi)\right],
\end{align}
where we have already integrated over outgoing angles $d\hat n$ using Eqs.~(\ref{eq:deltavchi}) and (\ref{eq:sigmabar}).

We perform this integral in Appendix \ref{App:heat} and find
\be
\dfrac{dQ_b}{dt} = \dfrac{2 m_b \rho_\chi \sigma_0  e^{-\frac{r^2}{2}}(T_\chi-T_b)}{(\mdm+m_b)^2\sqrt{2\pi} u_{\rm th}^3}  + \dfrac{\rho_\chi}{\rho_m}\dfrac{\mdm m_b}{\mdm+m_b} \Vdm D(\Vdm).
\label{eq:heat}
\ee
The first term, in the $r\to0$ limit, was derived in \cite{1311.2937,1408.2571}, but here we also find the second term, which is non-zero for $r\neq 0$.

By symmetry, $\dot Q_\chi$ is obtained by simply substituting $b\leftrightarrow \chi$ in Eq.~(\ref{eq:heat}). 
We see that these expressions, with the drag $D(\Vdm)$ in Eq.~(\ref{eq:drag}), conserve the total kinetic energy density in the baryon-DM fluid, i.e.
\be
n_\chi \dfrac{dQ_\chi}{dt} +n_b \dfrac{dQ_b}{dt} - \dfrac{\rho_\chi \rho_b}{\rho_m} D(\Vdm) \Vdm = 0.
\ee

Now that we know how the interactions change the energy of the baryons and DM at any given time, let us find how their temperatures are modified.

\subsection{Temperature evolution}
\label{sec:temp}

Using the expressions for the drag $D(\Vdm)$, in Eq.~(\ref{eq:drag}), and the heating rates $\dot Q_b$ and $\dot Q_\chi$, in Eq.~(\ref{eq:heat}), we can write the equations of the temperature evolution, \cite{1408.2571,1311.2937}. In our analysis we also evolve the relative velocity $\Vdm$. The set of equations we will have to solve is then
\begin{align}
\dfrac{d T_\chi}{da} &= -2 \dfrac{T_\chi}{a} + \dfrac {2\dot Q_\chi}{3 a H}  ,
\label{eq:Tchi}
\\
\dfrac{d T_b}{da } &= -2\dfrac{T_b}{a} + \dfrac {\Gamma_C}{a H} (T_\gamma-T_b) + \dfrac {2 \dot Q_b}{3 a H}, 
\label{eq:Tgas}
\\
\dfrac{d \Vdm }{da} &= -\dfrac{\Vdm}{a} -  \dfrac{D(\Vdm)}{a H} ,
\label{eq:vrel}
\end{align}
where we have assumed the photon temperature $T_\gamma$ is unaltered, $H$ is the Hubble parameter and $\Gamma_C$ is the Compton interaction rate,
which depends on the free-electron density $n_e$. Since the free-electron abundance also depends on the baryon temperature through the recombination rate, we must solve for Eqs.~(\ref{eq:Tchi})-(\ref{eq:vrel}) simultaneously with the free-electron fraction $x_e = n_e/n_H$
\be
\dfrac{d x_e}{da } = - \dfrac{C}{aH} \left( n_H \mathcal A_B x_e^2 - 4 (1-x_e) \mathcal B_B e^{3E_0/(4T_\gamma)} \right),
\label{eq:xe}
\ee
where $C$ is the Peebles factor \cite{Peebles:1968ja}, $E_0$ is the ground energy of Hydrogen, and $\mathcal{A}_B(T_b, T_\gamma)$ and $\mathcal{B}_B(T_\gamma)$ are the effective recombination coefficient and the effective photoionization rate to and from the excited state respectively \cite{1006.1355,1011.3758}.

For convenience, we parametrize the results in terms of a dimensionless cross section $\sigma_{41}$, defined as
\be
\sigma_{41} \equiv \dfrac{\sigma_0}{ 10^{-41}\rm cm^2},
\ee
so that $\sigma_{41}\leq 3.2 (\mdm/ \rm GeV)$ is the 95\% C.L. constraint from CMB-analysis \cite{1311.2937}, valid only for $\mdm\gg m_b$.

\subsection{Limiting cases}

To gain understanding of the implications of Eq.~(\ref{eq:heat}) it is enlightening to study the
extreme cases of very-heavy and very-light dark matter.

$\bullet$ For very massive dark matter ($\mdm\gg m_b \approx 1$ GeV), the first term in Eq.~(\ref{eq:heat}) is small and the second one dominates, which means that the new effect we have calculated is more relevant than the previously-known result.
In this limit we then have $\dot Q_b = (\rho_\chi/\rho_m) m_b \Vdm D(\Vdm) [1+O(m_b/\mdm)]$, which means $\dot Q_b \propto \sigma_0/\mdm$. 
Equivalently, the DM heating term will be given by  $\dot Q_\chi = (\rho_b/\rho_m) m_b \Vdm D(\Vdm) [1+O(m_b/\mdm)]$, so that $\dot Q_\chi \propto \sigma_0/\mdm$ as well, so for $\mdm\gg m_b$ the constraints we will find will behave as $\sigma_0 \propto \mdm$.

$\bullet$ In the opposite limit, in which $\mdm \ll m_b$, we find that the temperature-independent heating term (second term in Eq.~(\ref{eq:heat})) is linear in $\mdm$ and hence subdominant. The first term is roughly constant. Although $u_{\rm th}$ depends on $T_\chi/\mdm$, $T_\chi$ starts as zero and does not change unless there are interactions. This leads to a net mass-independent cooling $\dot{Q}_b < 0$, whereas the dark matter decouples, since $\dot Q_\chi \propto \mdm \to 0$.

Let us now briefly discuss the two limiting cases where either thermal or relative velocities dominate,

$\bullet$ When $\Vdm\ll u_{\rm th} \equiv \sqrt{T_\chi/\mdm + T_b/m_b}$ (thermal velocity dominates), we recover the results of Ref.~\cite{1408.2571}, where baryons get cooled down and tend to thermalize with the dark matter fluid. This is shown in Fig.~\ref{fig:TbTchi} as the ``$V_{\chi b, 0}=0$" case.

$\bullet$ In the limit where $\Vdm$ is much bigger than $u_{\rm th}$, the second term in Eq.~(\ref{eq:heat}) dominates, which causes a net heating of the baryon fluid. However, the overall rate of interactions (and hence net heating or cooling) is suppressed for large velocities, due to the fact that the cross section is proportional to $v^{-4}$.

\subsection{Numerical results}

We solve the system Eqs.~(\ref{eq:Tchi})-(\ref{eq:xe}) for different values of $\sigma_{41}$ and $\mdm$, starting at $z = 1010$ 
with the baryons tightly coupled to the photon fluid  ($T_b=T_\gamma$) and with perfectly cold dark matter ($T_\chi=0$), although we tested that having slightly warm dark matter at recombination does not change our results significantly. We use cosmological parameters consistent with their current best-fit values \cite{1502.01589}. 
We have also checked that, for the values of $\sigma_{41}$ considered in our analysis, the system is not already tightly coupled 
at $z=1010$, which would require us to start evolving the system at an earlier redshift.

As for the initial conditions for $\Vdm$, we will solve the system for an array of values from zero initial velocity to three times the width of its Gaussian distribution. For purposes of illustration we will plot two different cases, one in which $V_{\chi b,0}= V_{\rm rms} = 29 $ km/s at initial redshift, and another in which $V_{\chi b,0}=0$, to show how the relative velocity affects the results. In the case with $V_{\chi b,0}\neq 0$, higher values of $\mdm$ imply a more significant heating of the baryons.

\begin{figure}[htbp!]

\includegraphics[width=1.0\linewidth]{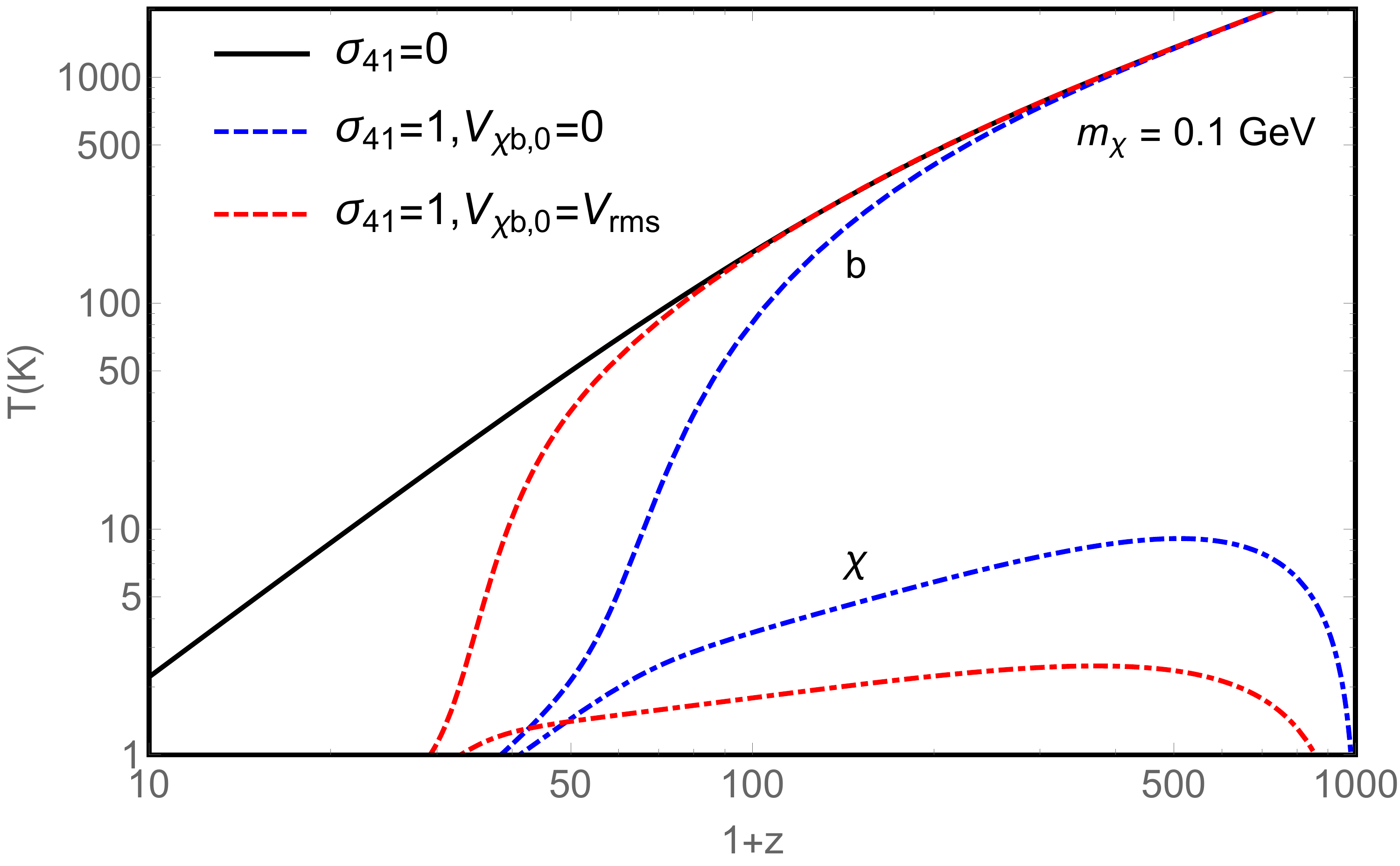}

\includegraphics[width=1.0\linewidth]{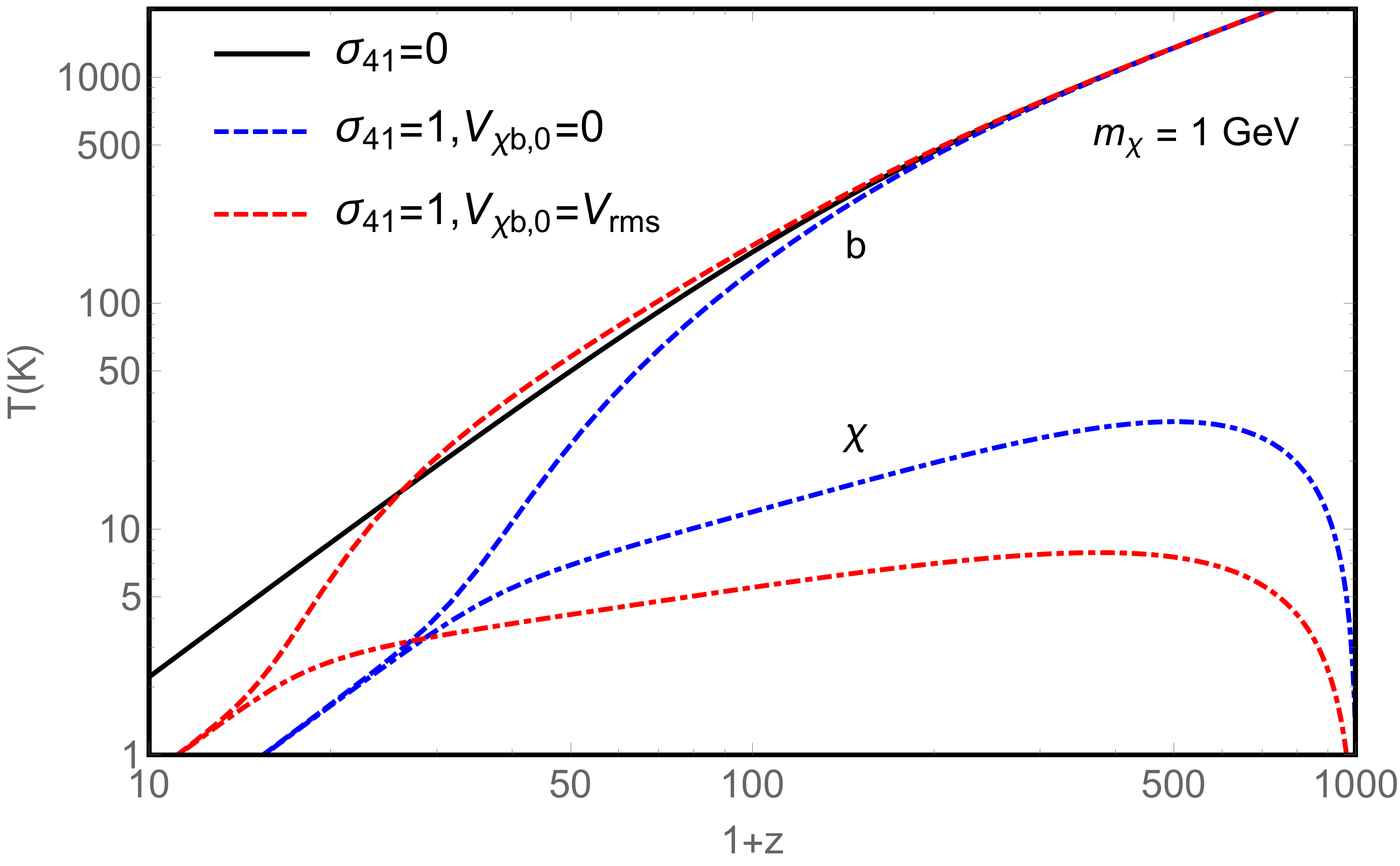}

\includegraphics[width=1.0\linewidth]{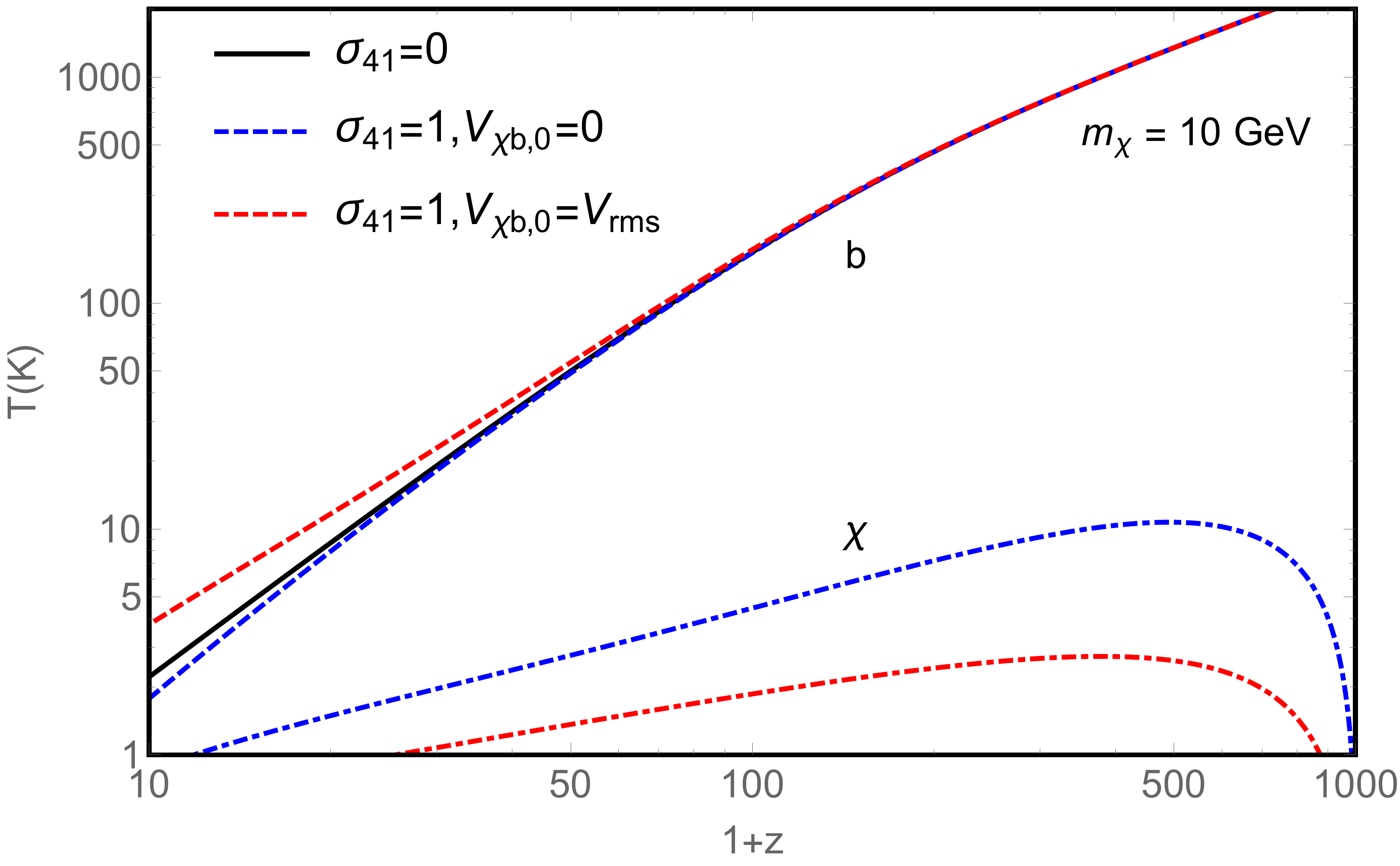}

\caption{Baryon temperatures (three upper curves) without interactions (solid curve) and when adding interactions with $\sigma_{41}=1$ (dashed-blue curve for the case where $V_{\chi b,0} = 0$ and red curve for $V_{\chi b,0} = V_{\rm rms}$), as well as dark-matter temperatures (two lower curves, dash---dotted-blue curve for the case where $V_{\chi b,0} = 0$ and red curve for $V_{\chi b,0} = V_{\rm rms} $). From top to bottom we show the results for $\mdm=0.1$, 1, and 10 GeV.} 
\label{fig:TbTchi}
\end{figure}

In Fig.~\ref{fig:TbTchi} we show how the baryon temperature changes with the strength of the interactions.
In the central and bottom panels we have $\mdm \geq m_p$. In those two figures it is explicit that having $V_{\chi b,0} \neq 0$ (red lines) induces extra heat in the system as a result of the damping of the relative velocity, which increases the temperature of both baryons and dark matter. However, when considering the case with $V_{\chi b,0} = 0$ (blue lines), the interactions cool down the baryons and only heat up the dark matter.
In the upper panel of Fig.~\ref{fig:TbTchi} we have set $\mdm = 0.1$ GeV. In this case it is clear that introducing interactions can only cool down the baryons, albeit with a more pronounced temperature drop in the  $V_{\chi b,0} = 0$ case.

\section{Effects on the 21-cm dark ages signal}
\label{sec:21cm}

We have seen how the baryon and dark matter temperatures change when adding interactions. Now we will study how this modified baryon temperature gives rise to a different spin temperature for the gas during the dark ages, which in turn modifies the 21-cm brightness temperature we would observe.

\subsection{21-cm brightness temperature}

The electronic ground state of neutral Hydrogen is split into two hyperfine states, a singlet spin-0 and a triplet spin-1 configuration. The singlet state has a smaller energy, with the transition from the triplet to the singlet corresponding to a wavelength of 21 cm. Because of its very long wavelength it is hard to confuse with any other redshifted line, making it a very unique probe of the physics of the early universe \cite{astro-ph/0312134}.

We define the spin temperature of the baryon gas through the ratio of the populations of the triplet to the singlet states,
\be
\dfrac{n_1}{n_0} = 3 e^{-T_*/T_s},
\ee
where $T_*=0.068\, K = 5.9\,\mu$eV is the energy corresponding to the 21-cm transition. During the dark ages the upward and downward transitions are much faster than the evolution of the universe. This means that we can use the quasi-steady-state approximation and, to a good accuracy, find the values of $n_1$ and $n_0$ for which there is equilibrium, 
\be
n_0 \left( C_{01} + R_{01} \right)=n_1 \left( C_{10} + R_{10} \right),
\ee
where $R_{ij}$ are the rates of radiative transitions of the CMB blackbody photons and $C_{ij}$ are the collisional transition rates \cite{astro-ph/0702600}. We will always have $T_* \ll T_b, T_\gamma$, in which case the spin temperature is very well approximated by
\be
T_s = T_\gamma + \dfrac{C_{10}}{C_{10} + A_{10} \frac{T_b}{T_*}},
\ee
where $A_{10}$ is the downward spontaneous Einstein coefficient of the 21-cm transition. We neglect the  Wouthuysen-Field effect \cite{Wouthuysen,Field,astro-ph/0507102}
  that would arise from inelastic scattering of Lyman-$\alpha$ photons after the first stars are created. 

The fact that the spin temperature is smaller than the CMB temperature during the dark ages enables the Hydrogen gas to resonantly absorb more CMB photons with a rest wavelength of 21 cm than it emits. We would observe this effect as a smaller brightness temperature of the CMB. Let us define the temperature ($T_{21}$) of the 21-cm line (redshifted to today) as the brightness temperature contrast with respect to the CMB, which is given by
\be
T_{21} = \dfrac{T_s-T_\gamma}{1+z} \left( 1-e^{-\tau}\right) \approx \dfrac{T_s-T_\gamma}{1+z} \tau,
\label{eq:T21}
\ee
where we have used the fact that the optical depth $\tau$ is very small, and is in fact given by
\be
\tau = \frac{3}{32 \pi} \frac{T_*}{T_s} n_{\rm HI} \lambda_{21}^3 \frac{A_{10}}{H(z) + (1+z) \partial_r v_r},
\label{eq:tau}
\ee
where $H(z)$ is the Hubble rate, $n_{\rm HI}$ is the density of neutral Hydrogen, $\lambda_{21}\approx 21$ cm and $\partial_r v_r$ is the proper gradient of the peculiar velocity along the radial direction.

In the usual scenario the spin temperature follows the gas temperature as of decoupling and until $z\sim 30$, which makes it different from the CMB temperature in the redshift range $z\sim 30$ to 200. This creates a non-zero 21-cm line temperature $T_{21}$ in this range. As we have shown, dark-matter--baryon interactions can either cool down or heat up the baryons, thus changing the spin temperature. 

\begin{figure}[htbp!]

\includegraphics[width=1.0\linewidth]{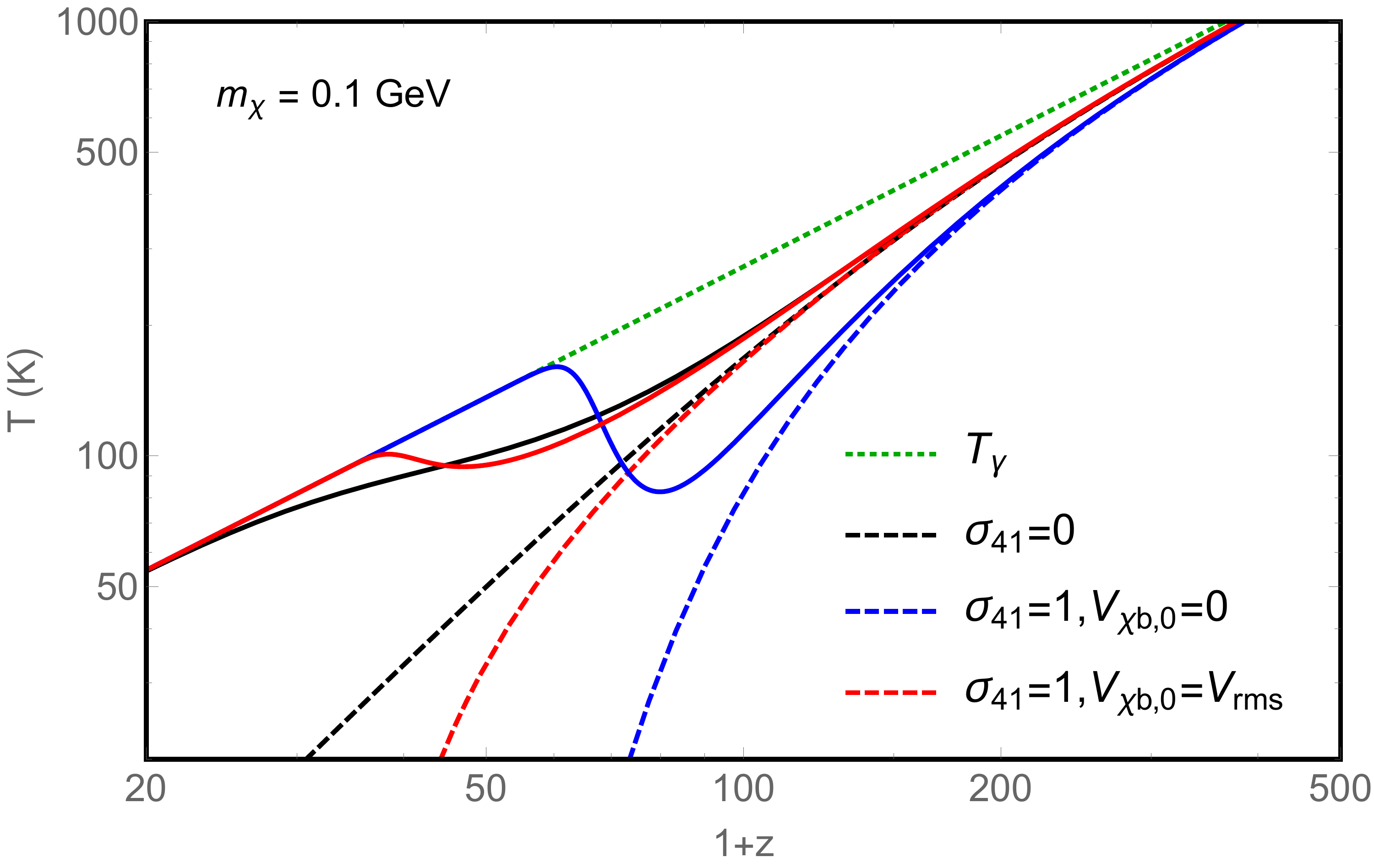}

\includegraphics[width=1.0\linewidth]{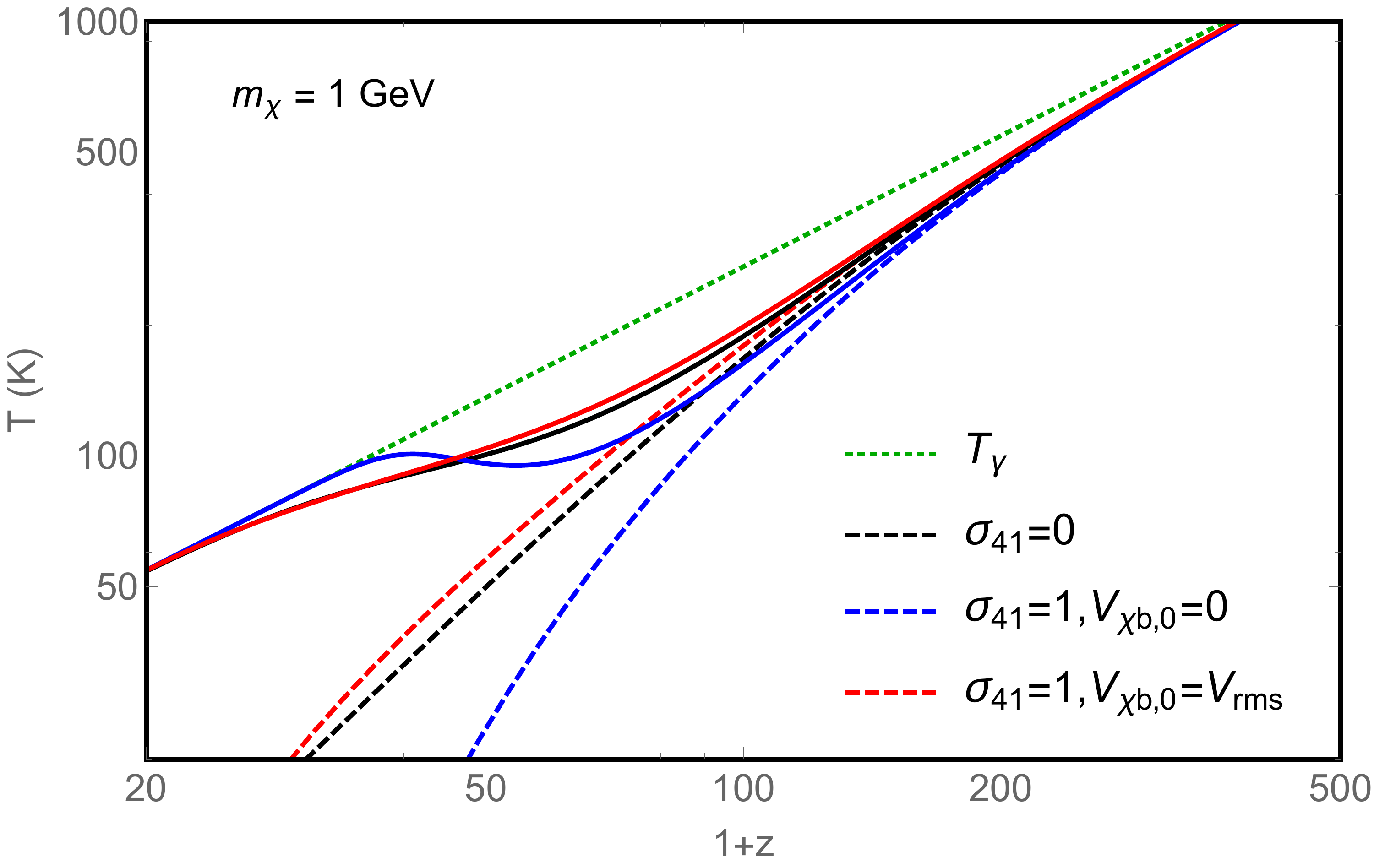}

\includegraphics[width=1.0\linewidth]{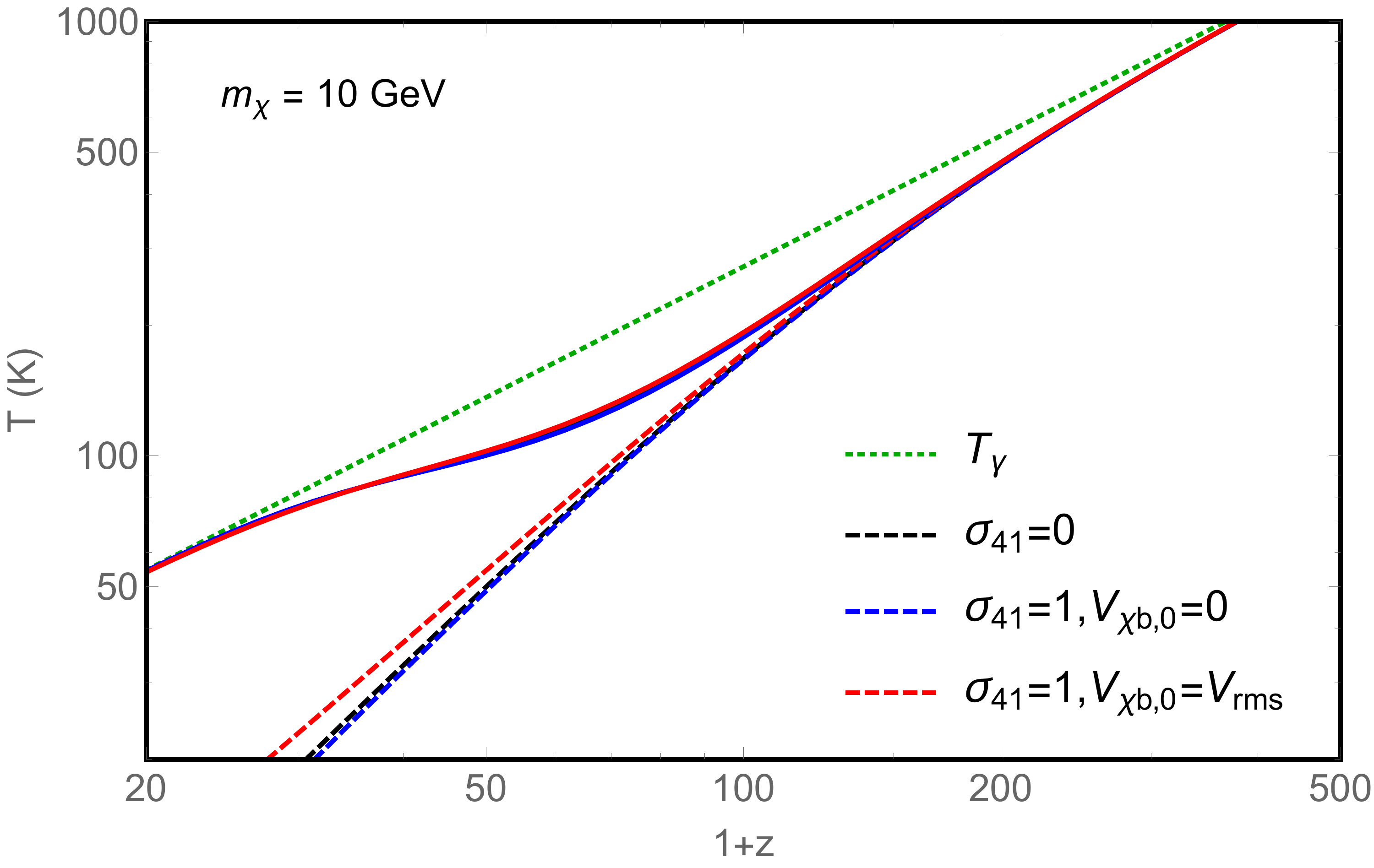}

\caption{Values of the spin temperature (dashed curves) and the gas temperature (solid curves) for the collisionless case (black curve) and when including collisions (blue curve for $V_{\chi b,0}=0$ and red curve for $V_{\chi b,0}=V_{\rm rms}$), as well as the CMB temperature in the dashed-green curve. From top to bottom we show the results for $\mdm=0.1$, 1, and 10 GeV.}
\label{fig:Tspin}
\end{figure}

We show this effect in Fig.~\ref{fig:Tspin}, where we plot for reference the CMB temperature, as well as the usual non-interacting gas and spin temperatures. We also plot the gas and spin temperature for interacting cases with either $V_{\chi b,0}=0$ or $V_{\chi b,0}=V_{\rm rms}$. The deviation of the spin temperature in the interacting cases (blue and red curves) is apparent, even for a cross section of $\sigma_{41}=1$, compatible with CMB bounds.

\begin{figure}[htbp!]

\includegraphics[width=1.0\linewidth]{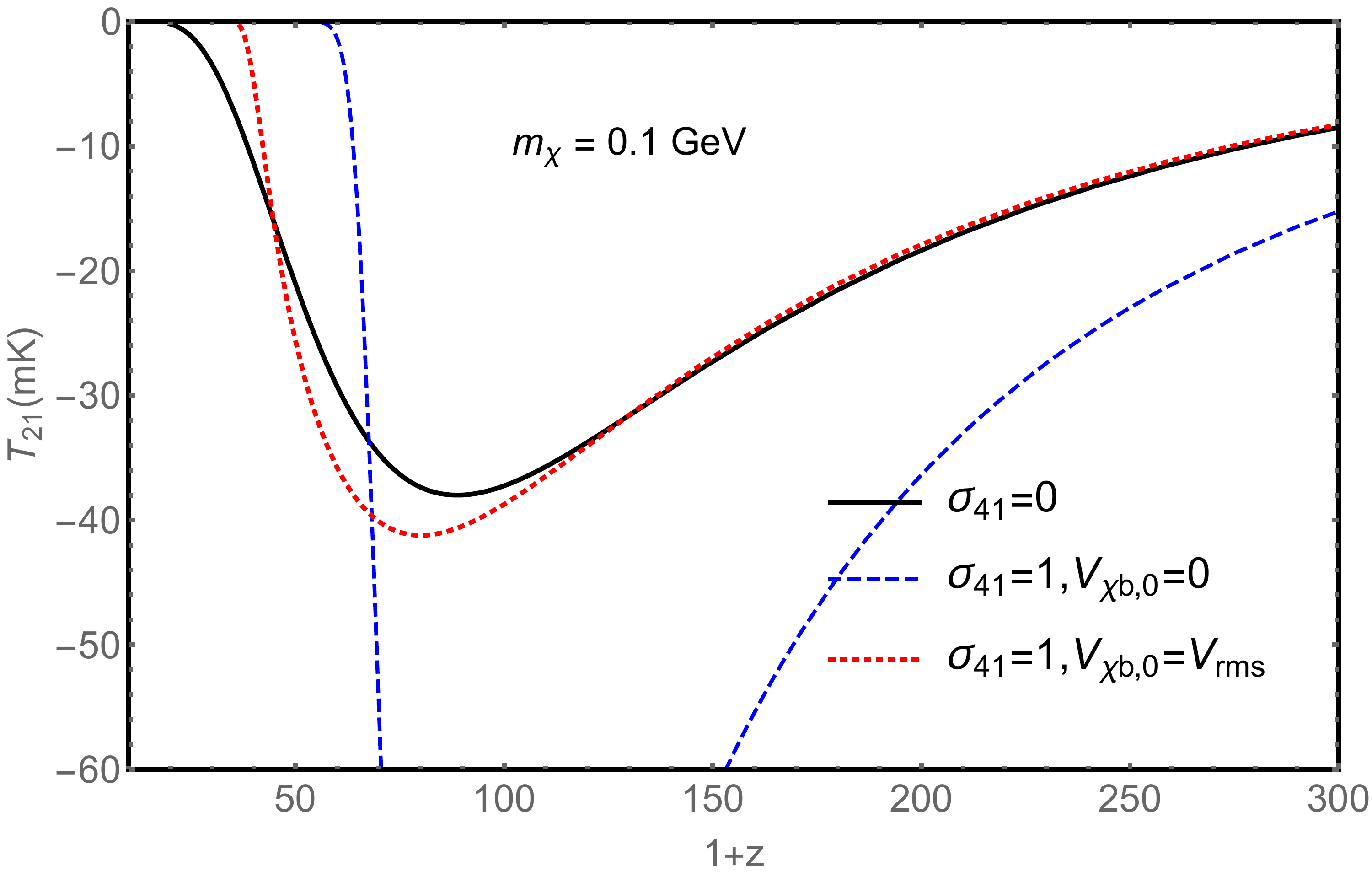}

\includegraphics[width=1.0\linewidth]{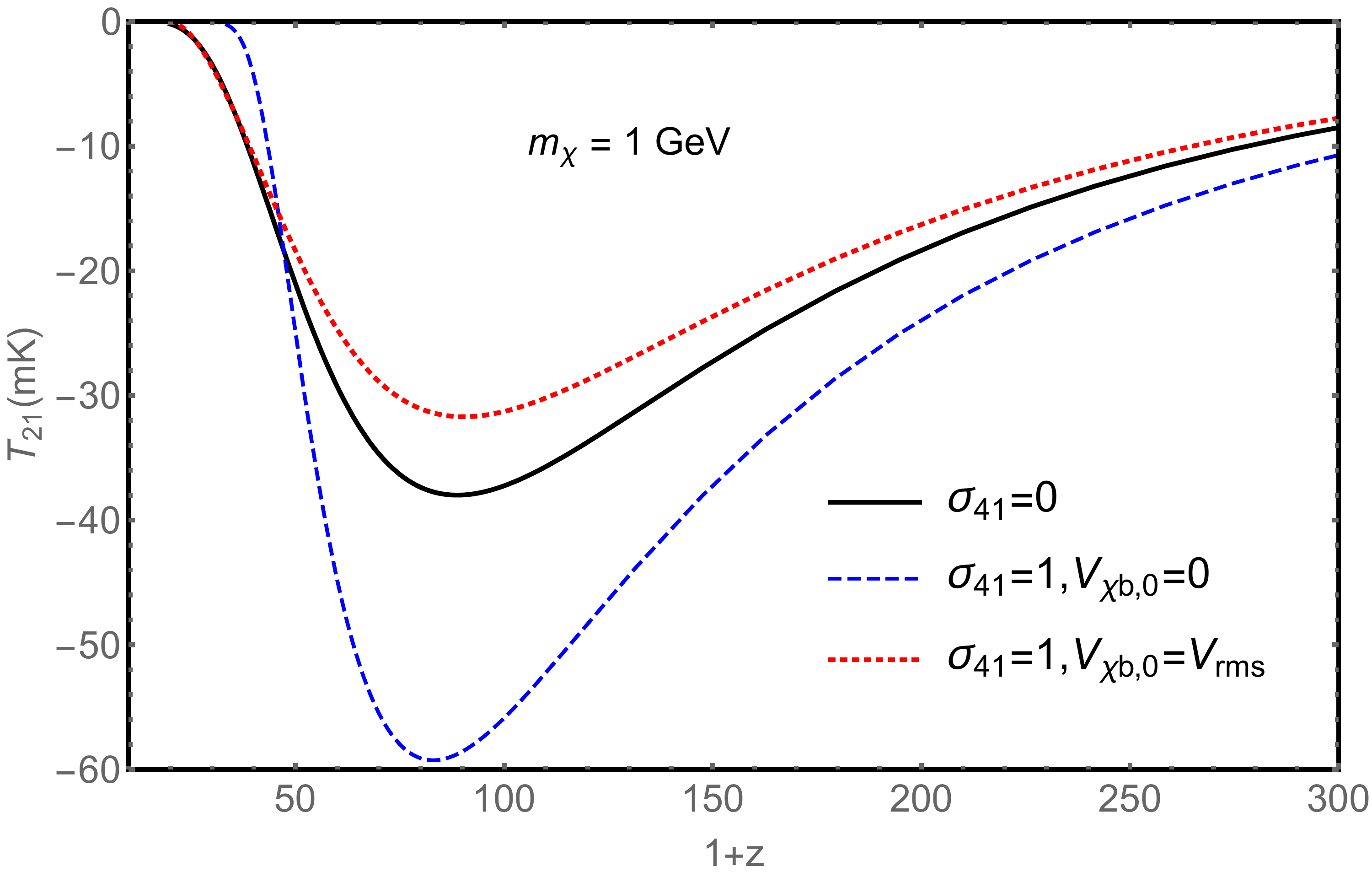}

\includegraphics[width=1.0\linewidth]{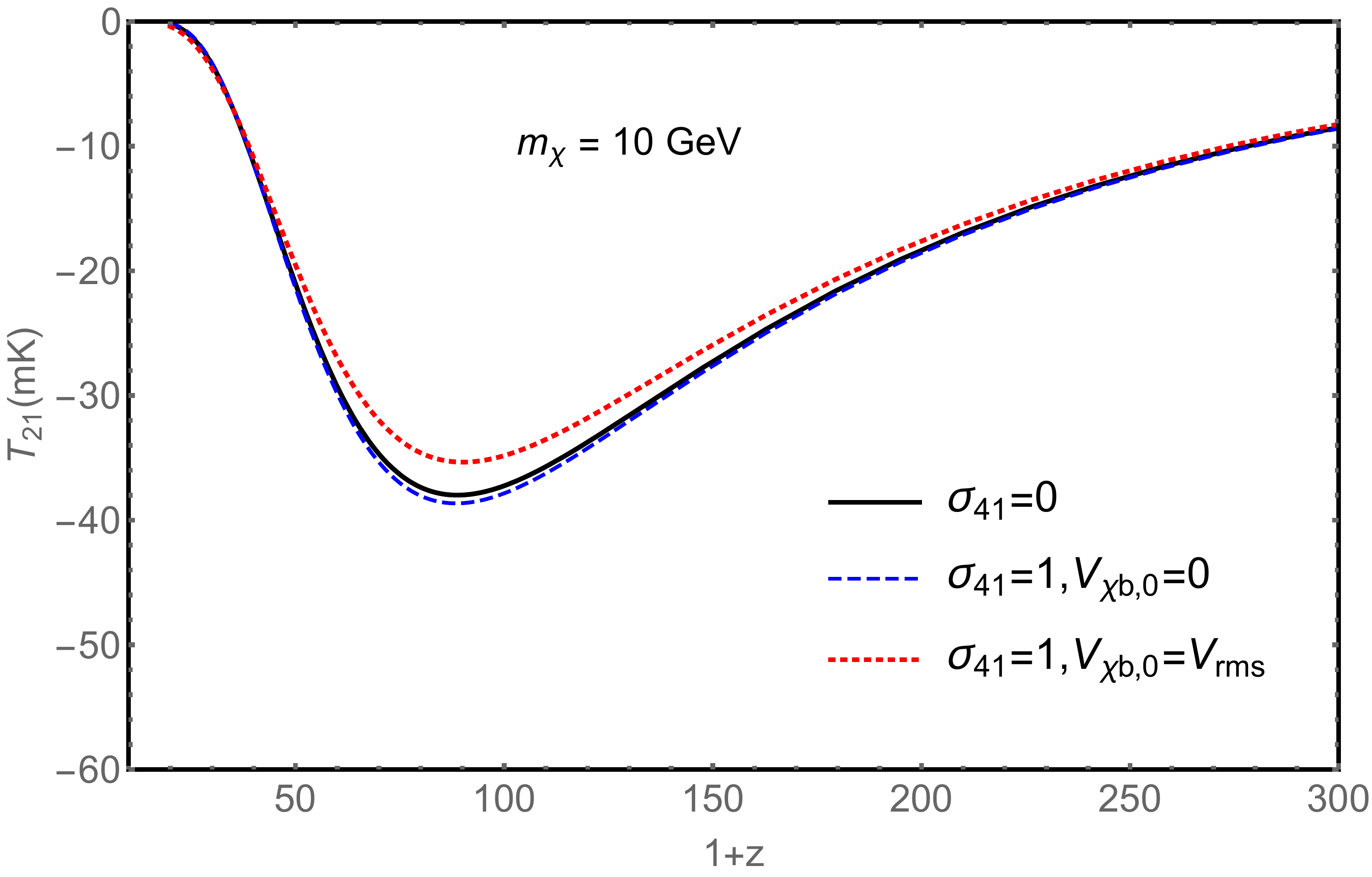}

\caption{Values of the average brightness temperature of the 21-cm line for the collisionless case (solid-black curve), the case with interactions (blue-dashed curve for $V_{\chi b,0}=0$, purple--dot-dashed curve for $V_{\chi b,0}=V_{\rm rms}$), and the average over initial velocities in the red-dotted curve. From top to bottom we show the results for $\mdm=0.1$, 1, and 10 GeV.} 
\label{fig:Tbr}
\end{figure}

If there is more heating than cooling of the baryons, the 21-cm brightness temperature decreases in magnitude, since the spin temperature is closer to the CMB temperature during the dark ages. Cooling of the baryons increases the brightness temperature, as long as the spin temperature stays coupled to the baryons. In Fig.~\ref{fig:Tbr} we plot the 21-cm brightness temperature, from Eq.~(\ref{eq:T21}),
for different values of the relative velocity and DM mass.
It is interesting to note that the heating increases with the mass of the dark matter, as predicted, so that the average brightness temperature $\bar T_{21}$ during the dark ages is higher when including interactions.

\subsection{Global signal}

\begin{figure}[htbp!]
\hspace{-1.cm}
\includegraphics[width=1.\linewidth]{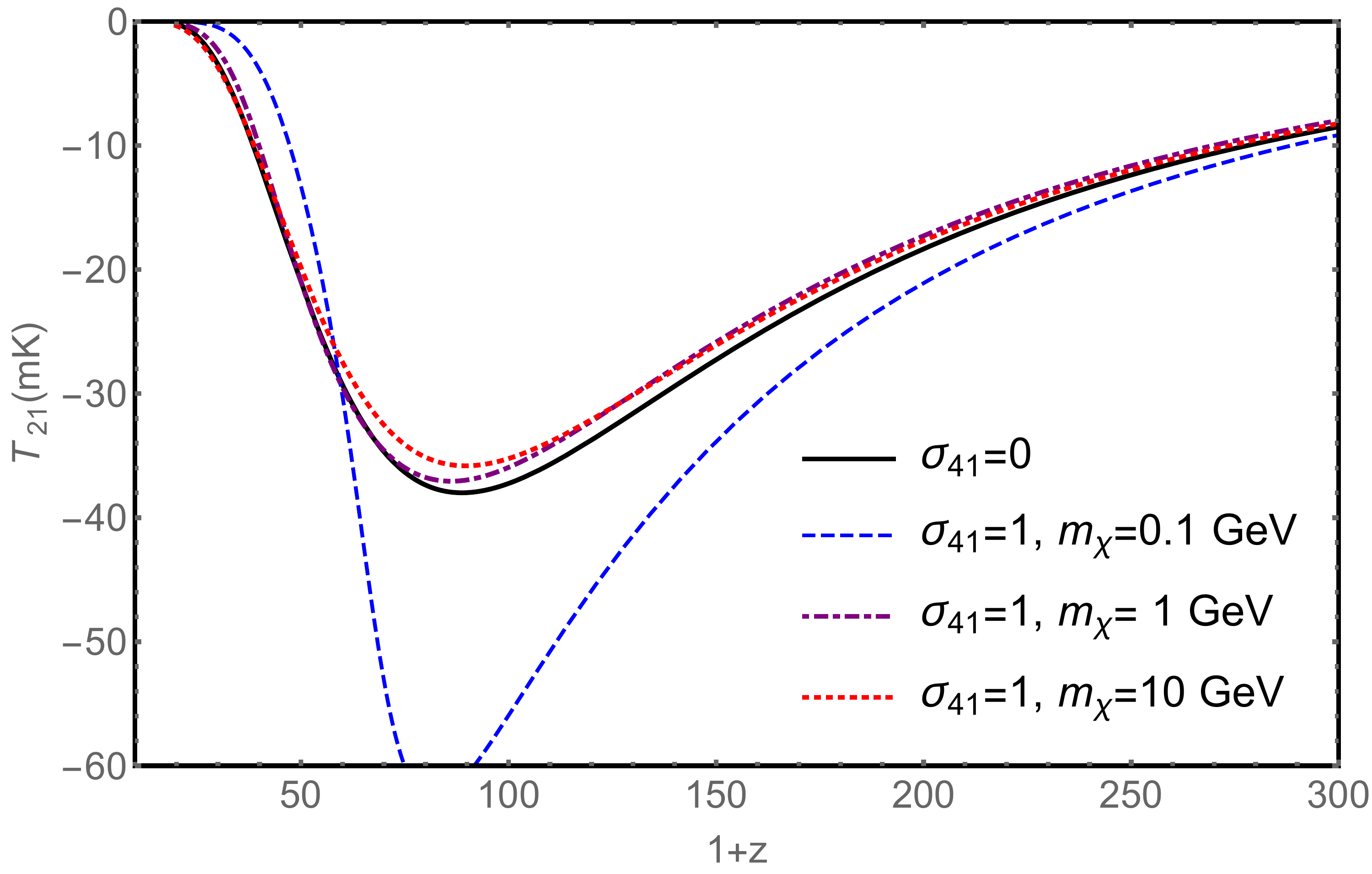}
\caption{Values of the brightness temperature of the 21-cm line for the collisionless case in solid-black curve, and three with interactions ($\sigma_{41}=1$), in dashed-blue curve $\mdm=0.1$ GeV, in dot-dashed---purple $\mdm=1$ GeV, and in dotted-red curve $\mdm=$ 10 GeV.} 
\label{fig:global}
\end{figure}

Let us define $\bar T_{21}(\Vdm)$ as the  brightness temperature in the absence of density perturbations. In the standard scenario this quantity is spatially homogeneous and 
is termed the global 21-cm signal. Once DM-baryon interactions are included, $\bar T_{21}(\Vdm)$
 is still a function of the initial relative velocities. We calculate its average
over said initial velocities as 
\be
\VEV{\bar T_{21}} = \int d^3  V_{\chi b,0} \bar T_{21}(V_{\chi b,0}) \mathcal P(V_{\chi b,0}),
\label{eq:avg}
\ee
with the probability distribution $\mathcal P(V_{\chi b,0})$ given by Eq.~(\ref{eq:pv}). 
We show this quantity in Fig.~\ref{fig:global} for the interacting case and for three different DM masses.

\subsection{21-cm fluctuations}

As we have shown, the brightness temperature $T_{21}$ of the 21-cm line is modified by the inclusion of interactions, and this modification depends on the initial relative velocity. 
The large-scale fluctuations of the relative velocity will therefore be imprinted on the brightness temperature, since two regions with different initial relative velocities will appear with different brightness temperatures (compare blue and red lines in Fig.~\ref{fig:Tbr}), which will actually generate an additional contribution to the power spectrum of the 21-cm fluctuations. Let us calculate it.

The standard deviation of $T_{21}$ as a function of $V_{\chi b,0}$ is
\be
T_{21,\rm rms} \equiv \sqrt{\VEV{T_{21}^2}-\VEV{T_{21}}^2}.
\label{eq:defrms}
\ee

\begin{figure}[htbp!]
\hspace{-1. cm}
\includegraphics[width=1.\linewidth]{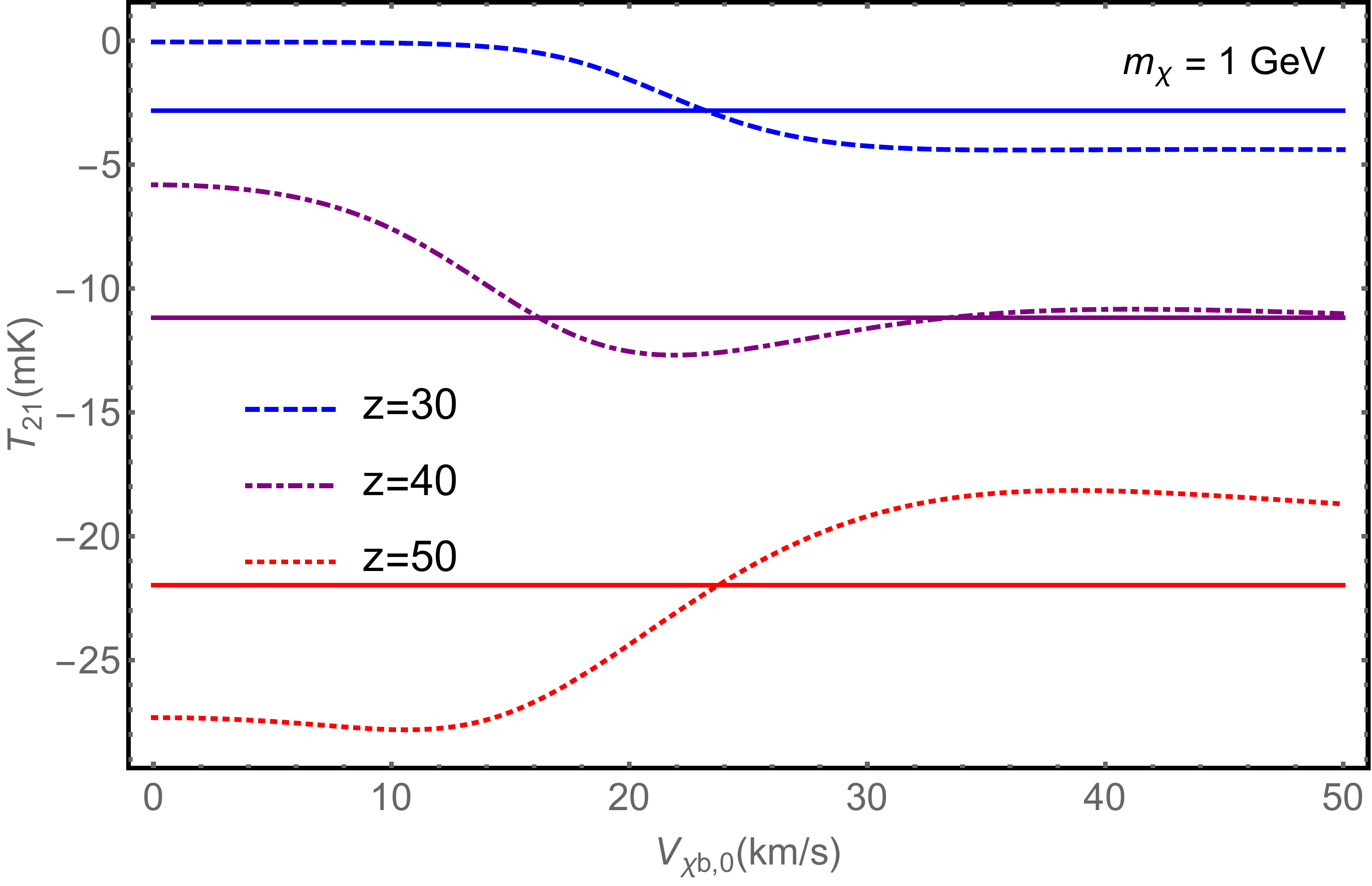}
\caption{Brightness temperature $\bar T_{21}$ of the 21-cm line for $\mdm=1$ GeV and $\sigma_{41}=1$ at redshifts $z=30$ in dashed-blue curve, $z=40$ in dash-dotted--purple curve and $z=50$ in dotted-red. We also show the average over velocities for each redshift, as defined in Eq.~(\ref{eq:avg}), in solid curves and their corresponding colors.} 
\label{fig:T21v}
\end{figure}

\begin{figure}[htbp!]
\hspace{-1. cm}
\includegraphics[width=1.\linewidth]{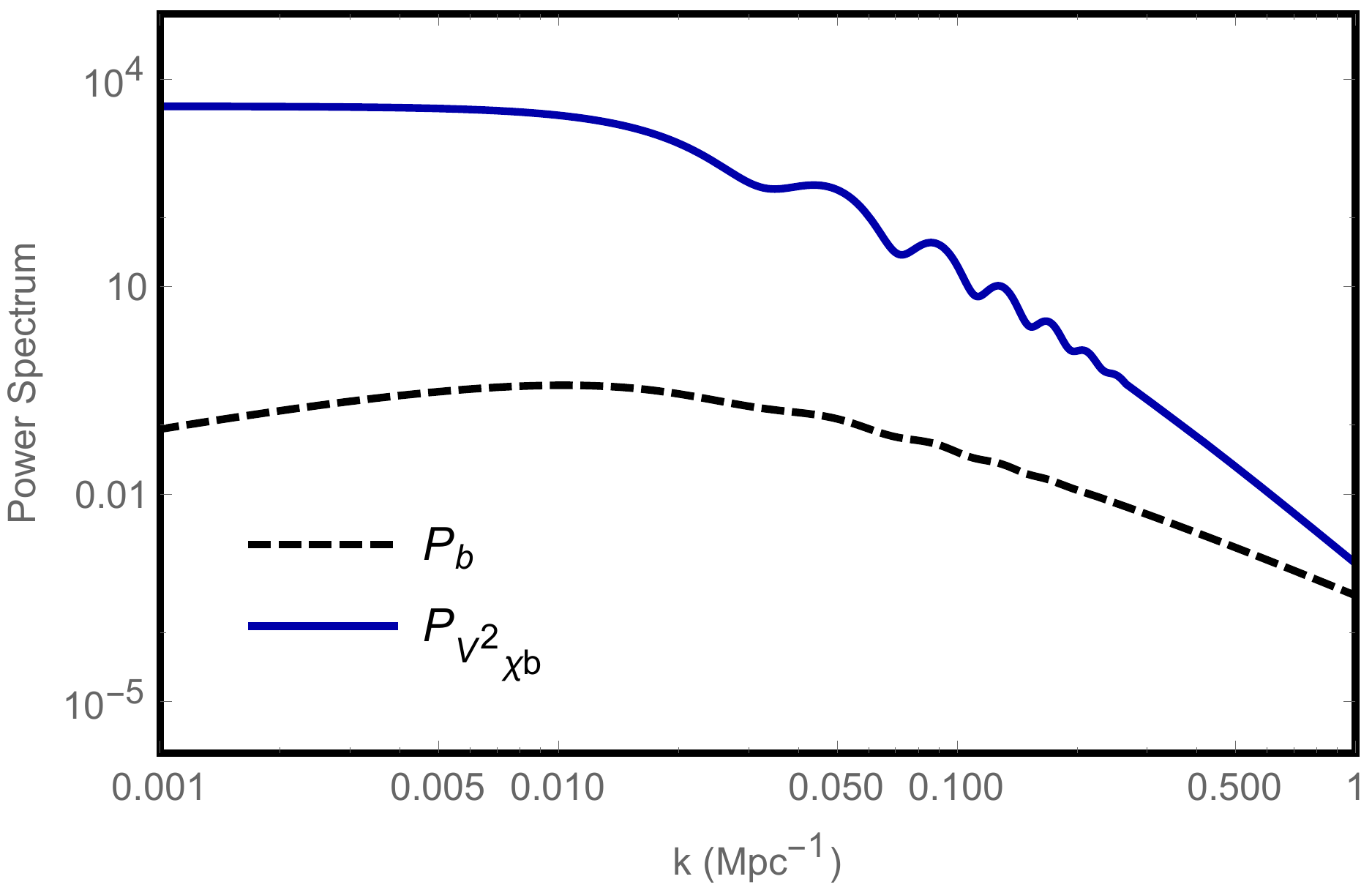}
\caption{Power spectra as a function of $k$. In solid-blue curve we show the power spectrum of $\Vdm^2$, as defined in Eq.~(\ref{eq:vPowSp}), and in dashed-black curve the standard power spectrum of baryon overdensities at redshift $z=30$.} 
\label{fig:Deltasq}
\end{figure}

Even if $T_{21}$ had no explicit spatial dependence, it would fluctuate because relative velocities are not homogeneous. 
In principle, to compute the power spectrum of $T_{21}$, one should first compute its two-point correlation function. 
This is obtained by integrating over the six-dimensional joint probability distribution of the relative velocities at two different points (see  Ref.~\cite{1312.4948}).
To simplify matters we shall make the following approximation
\be
T_{21}(V_{\chi b,0}) \approx \VEV{T_{21}} +  T_{21,\rm rms} \sqrt{\dfrac 2 3}\left( 1 -\dfrac{\Vdm^2}{V_{\rm rms}^2} \right),
\ee
which has the advantage of resulting in simple analytic expressions \cite{1110.2182} while still reproducing adequately the variance of $T_{21}$. For illustration we show $T_{21}$ as a function of $V_{\chi b,0}$ for the $\mdm=1$ GeV case in Fig.~\ref{fig:T21v}.
We calculate the power spectrum of $T_{21}(V_{\chi b,0})$ in
this approximation to be
\be
\VEV{T_{21}(\mathbf k) T_{21}^*(\mathbf k')} =  T_{21,\rm rms}^2 P_{\Vdm^2}(k) (2\pi)^3 \delta_D(\mathbf k+\mathbf k'),
\label{eq:vPowSp}
\ee
where $P_{\Vdm^2}$ is the power spectrum of $\sqrt{2/3}(1-V_{\chi b, 0}^2/V_{\rm rms}^2)$. We plot $P_{\Vdm^2}(k)$ in Fig.~\ref{fig:Deltasq}.

Our observable, the brightness temperature of the 21-cm line, varies in space through its dependence on the baryon density $n_b$, as well as on the initial relative velocities $V_{\chi b,0}$. 
To linear order in density perturbations the temperature of the 21-cm line, Eq.~(\ref{eq:T21}), will be given by \cite{1506.04152}
\be
T_{21} = \bar T_{21}(V_{\chi b,0}) + \dfrac{dT_{21}}{d\delta} \delta,
\label{eq:T21fo}
\ee
where $\delta \equiv (n_b-\bar n_b)/\bar n_b$, $dT_{21}/d\delta$ is a well-known function of redshift for $V_{\chi b,0}=0$, and $\bar T_{21} = \bar \tau (\bar T_s-T_\gamma)/(1+z)$ is the unperturbed value of the brightness temperature. Both $\bar T_{21}$ and $dT_{21}/d\delta$ depend on the initial relative velocities. 
The average over initial velocities of $T_{21}$ is then
\be
\VEV{T_{21}} = \VEV{\bar T_{21}} + \VEV{\dfrac{dT_{21}}{d\delta}} \delta.
\ee
We can, however, approximate $\VEV{dT_{21}/d\delta}\approx dT_{21}/d\delta(V_{\chi b,0}=0)$, since the error made in the 21-cm temperature would be of order $\delta  T_{21}(k) \sim T_{21,\rm rms} \delta$, which is subdominant.
We can calculate the variance of $T_{21}$ over both initial relative velocities and overdensities (as in the usual power spectrum) to find
\ba
P_{T_{21}}(k)= \bar T_{21,\rm rms}^2 P_{\Vdm^2}(k) 
+ \left( \alpha(z) + \bar T_{21} \frac{k_{||}^2}{k^2} \right)^2 P_b(k,z),
\label{eq:Delta21}
\end{align}
where $k_{||}$ is the magnitude of $k$ in the line-of-sight direction, $\alpha(z)$ as defined in Ref.~\cite{1506.04152}, and $P_b$ is the usual baryon power spectrum.

We can convert easily from $k$-space to $\ell$-space by using a harmonic transform \cite{astro-ph/0001303}, which is exact in the case of the flat-sky limit and still a very good approximation for $\ell\geq 10$, which should be good enough for our order-of-magnitude estimates. We define the angular power spectrum for the standard fluctuations (std) as
\be
C_\ell^{\rm std} = \dfrac 1 {r^2} \int \dfrac{d k_{||}}{2\pi} \left | \tilde W (k_{||})\right |^2   \left( \alpha + \bar T_{21} \dfrac{k_{||}^2}{k^2} \right)^2 P_b\left(k\right),
\label{eq:harmtr1}
\ee
where $k\equiv \sqrt{\ell^2/r^2+k_{||}^2}$, and $\tilde W(k_{||})$ is the window function. The new angular power spectrum ($\Vdm$), due to interactions, will be
\be
C_\ell^{\Vdm} = \dfrac {\bar T_{21,\rm rms}^2} {r^2} \int \dfrac{d k_{||}}{2\pi} \left | \tilde W (k_{||})\right |^2  P_{\Vdm^2}\left(k\right).
\label{eq:harmtr2}
\ee

\begin{figure}[htbp!]

\includegraphics[width=1.0\linewidth]{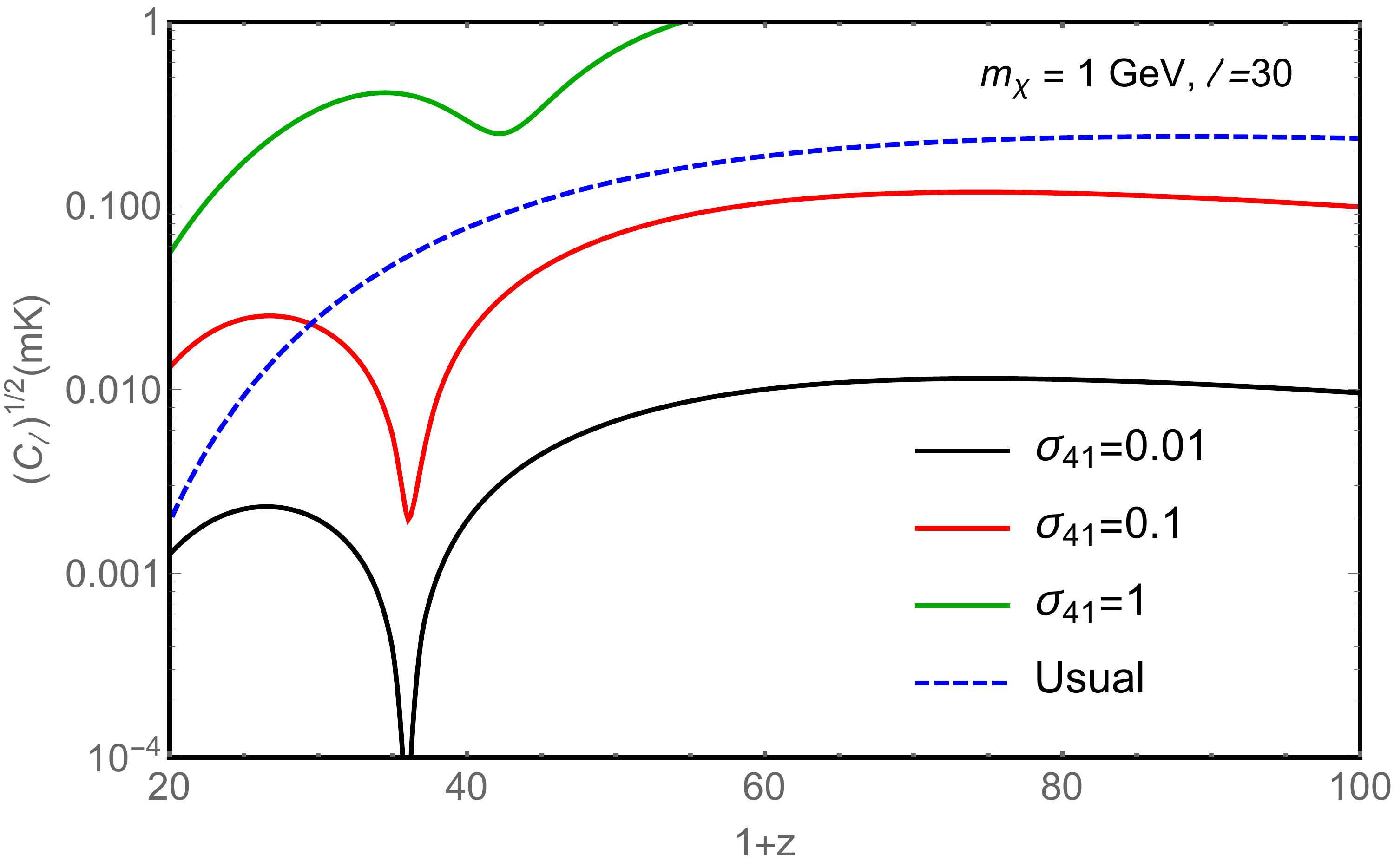}

\includegraphics[width=1.0\linewidth]{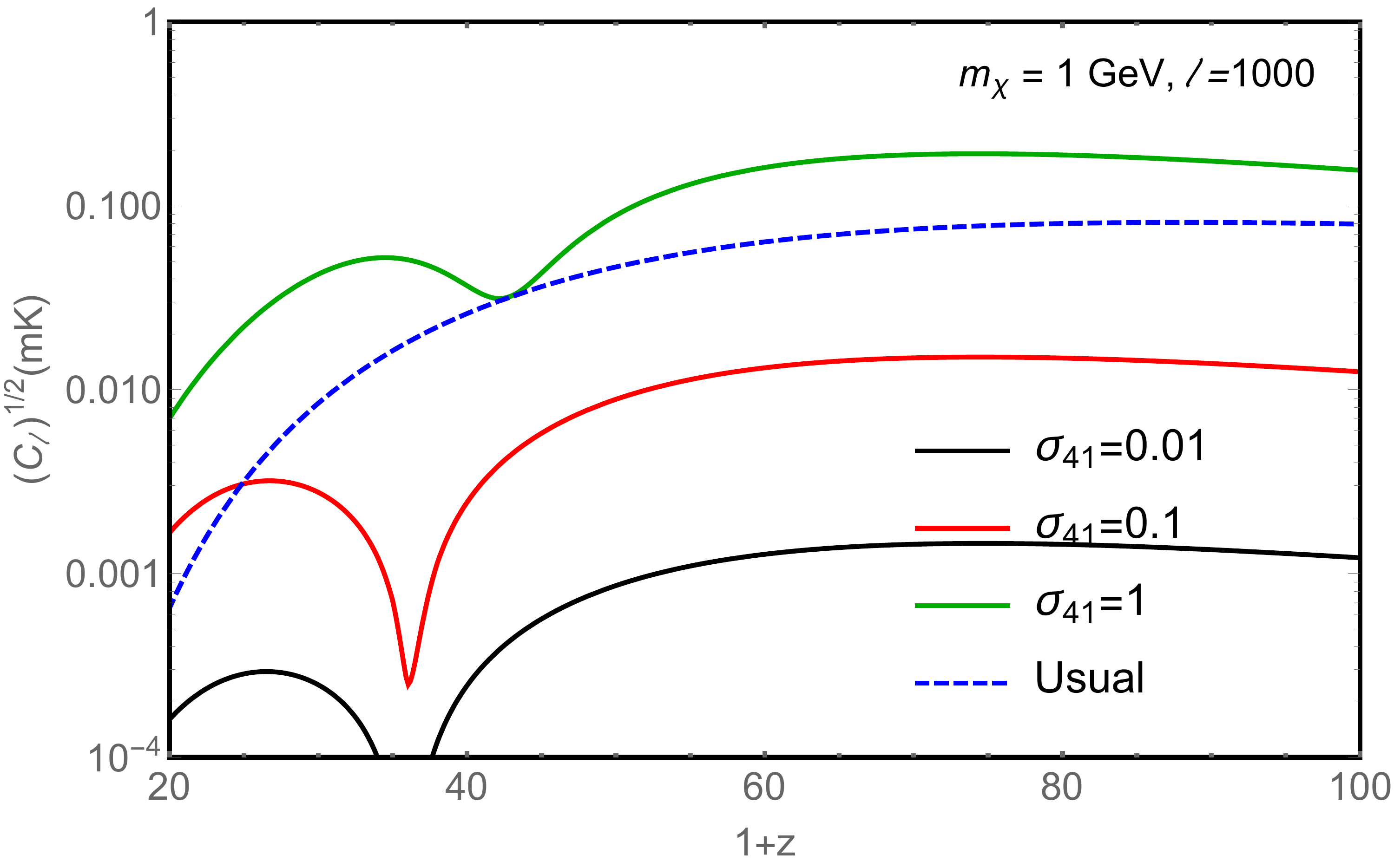}

\caption{Amplitude of the brightness-temperature fluctuations induced by relative velocity fluctuations for three different cross sections (solid-black curve for $\sigma_{41}=0.01$, red curve for $\sigma_{41}=0.1$, and green curve for $\sigma_{41}=1$), as well as for the usual baryon density perturbations. We calculate at two different scales: in the first panel we show the results for $\ell=30$ and in the second one for $\ell=1000$.} 
\label{fig:T21k}
\end{figure}

Before going into a full-scale analysis one might be interested in what would happen at a single $\ell$, and at different redshifts. 
We show in Fig.~\ref{fig:T21k} the value of the square root of the velocity power spectrum $(C^{\Vdm}_\ell)^{1/2}$ for different values of the cross section. For illustration purposes we also show the usual power spectrum $(C^{\rm std}_\ell)^{1/2}$, from Eq.~(\ref{eq:harmtr1}), where for simplicity we have taken $\alpha(z)/\alpha(z_0)(C^{\rm std}_\ell)^{1/2}(z_0)$ as a proxy for $(C^{\rm std}_\ell)^{1/2}(z)$ as well as a redshift-independent bandwidth of $\Delta \nu/\nu = 0.02$ to avoid recalculating the integral in Eq.~(\ref{eq:harmtr1}) for each redshift in this plot. We show the cases of $\ell=30$ and $\ell=1000$ in the upper and lower panels, respectively.

\section{Detectability}
\label{sec:detection}

So far we have shown that DM-baryons interactions
modify the baryon temperature, raising it or lowering it, depending on the initial relative velocity. Varying the baryon temperature will change the spin temperature and hence the brightness temperature of the 21-cm line. This quantity, also known as the ``global signal", is the main observable during the dark ages. We will study how to detect interactions with a global-signal experiment.

Moreover, since the temperatures depend on initial velocities, and these have a spatial-dependence, we have argued that there will be a new contribution to the power spectrum, which, at large scales, can overcome the standard one for values of the cross section of $\sigma_{41}\gtrsim 0.1$. We will study the detectability of this signal with interferometry later in this section.

\subsection{Global signal}

Let us start by analyzing the most direct effect of DM-baryon interactions, the change in the global signal during the dark ages. Next-generation experiments, such as 
NenuFAR, will survey the 21-cm line brightness temperature down to frequencies possibly as low as $\nu\sim 10$ MHz, which corresponds to a redshift $z > 100$. 

We have seen in Fig.~\ref{fig:Tbr} how the brightness temperature changes when adding interactions. We will use the amplitude of the brightness temperature at its peak as a proxy for the detectability of the signal, even though its very high redshift ($z\sim 90$) may make it unobservable. 

Let us first find the signal-to-noise ratio to detect interactions having a cross section $\sigma_{41}=1$. If we could determine the brightness temperature $\bar T_{21}$ at its peak with 5\% precision, we would be able to detect interactions with $\sigma_{41}=1$ at a signal-to-noise ratio $S/N\sim 10$ for $\mdm=0.1$ GeV,  $S/N\sim 0.5$ for $\mdm=1$ GeV, and $S/N\gtrsim 1$ for $\mdm=10$ GeV.

More interestingly, if we were able to improve the error by a factor of 5, reaching 1\% precision of peak-temperature determination, we would be able to detect cross sections as small as $\sigma_{41}\lesssim 0.04$ for $\mdm=0.1$ GeV, $\sigma_{41}\lesssim 0.1$ for $\mdm=1$ GeV, and $\sigma_{41}\lesssim 0.2$ for $\mdm=10$ GeV, all of which are beyond what can be achieved by current CMB analysis \cite{1311.2937}.

\subsection{Fluctuations}

We now turn our focus to the 
measurement of the 21-cm power spectrum, Eq.~(\ref{eq:Delta21}). In a
maximum-likelihood analysis, the Fisher forecast for the error
in the measurement of the amplitude $A$ of a power spectrum
$C_\ell$ is given by \cite{Jaffe:2000yt}\footnote{We assume that 
the likelihood function is Gaussian in the vicinity of its maximum
\cite{Jungman:1995bz,Zhao:2008re}.},
\be\label{eq:smallamp}
     \frac{1}{\sigma_A^2} = \sum_\ell \left( \frac{ \partial C_\ell }{ \partial A} \right)^2 \frac{1}{\sigma_\ell^2}.
\ee
For a given sky coverage $f_{\rm sky}$, the error for an individual 
$\ell$ in the estimated value $\widehat{A}$ is \cite{Jaffe:2000yt,Knox:2002pe,Kesden:2002ku} 
\be
  \sigma_\ell^{\widehat{A}} = \sqrt{\frac{2}{ f_{\rm
  sky}(2\ell+1)}}\left( C^{\rm std}_\ell +C^N_\ell\right),
\ee
where $C^N_\ell$ is the instrumental noise power spectrum, defined below in Eq.~(\ref{eq:NoisePS}),
and $C^{\rm std}_\ell$ is the standard power spectrum of 21-cm fluctuations (under the 
null hypothesis of no DM-baryon interactions), Eq.~(\ref{eq:Delta21}).

The minimum detectable amplitude $\widehat{A}$ at 1-$\sigma$ significance 
is thus
\be
     \sigma^{\widehat{A}}=\left[\frac{f_{\rm sky}}{2}
     \sum\limits_{\ell_{\rm min}}^{\ell_{\rm max}}
     \frac{(2\ell+1)(\tilde{C}^{\Vdm}_\ell)^2}{\left(
     C^{b}_{\ell}+C^N_{\ell}\right)^2}\right]^{-\frac{1}{2}},
\label{eq:estimatorerror}
\ee
where $\tilde{C}_{\ell}^{\Vdm}=C^{\Vdm}_\ell/A$ encodes the $\ell$ dependence of the velocity
power spectrum from DM-baryon interactions, and $\ell_{\rm min}=180/\theta$ is the 
largest scale accessible by an experiment with sky coverage $f_{\rm sky}=\theta^2$. Because of the use of the harmonic transform in Eq.~(\ref{eq:harmtr1}), we take $\ell_{\rm min}=15$. This should not affect the results significantly since there are very few modes at lower $\ell$.

We will consider two different scenarios, first a realistic experiment modeled after SKA that could be taking data within the next few years and second a more idealized experiment whose noise level will be low enough to detect the primordial power spectrum at redshift $z=30$ (but still not cosmic-variance limited, since the usual primordial power spectrum vanishes for smaller redshifts but the noise will not).

We will study the redshift range $z=20$ to 30, at the very end of the dark ages. This range is chosen to avoid complex astrophysical processes at low redshift as well as to still be observable from Earth. We may not be fully free of contamination, however, since the epoch of the formation of the first stars is unknown, and the X-rays generated during star formation may start to heat up the gas at $z\lesssim 25$ \cite{1212.0513}. Moreover, accreting intermediate-mass black holes (sometimes termed miniquasars) may also be an important source of X-rays during this era \cite{astro-ph/0304131,astro-ph/0506712,astro-ph/0608032}. 
Once data of the gas temperature during the dark ages are acquired, a careful analysis should take these processes into account along with the heating produced by DM-baryon interactions, and by studying their different redshift behaviors and angular structures, disentangle them. We motivate future work to address this issue.

The angular noise power spectrum of an interferometer is given by \cite{astro-ph/0311514}
\be
     \ell^2 C_\ell^N= \frac{(2\pi)^3 T_{\rm sys}^2(\nu) }{ \Delta \nu
     t_o  f_{\rm cover}^2} \left(\frac{\ell}{\ell_{\rm
     cover}(\nu)}\right)^2,
     \label{eq:NoisePS}
\ee
where $\ell_{\rm cover}(\nu)=2\pi D/\lambda$ is the maximum
multipole at frequency $\nu$ (corresponding to wavelength
$\lambda$) that can be measured with an array of dishes with
maximum baseline $D$,
covering a total area $A_{\rm total}$ with a covering fraction
$f_{\rm cover}\equiv N_{\rm dish} A_{\rm dish} /A_{\rm total}$,
in a frequency window $\Delta \nu$ with an observing time
$t_o$. The system temperature is given by $T_{\rm sys} 
\sim 180\left(\nu/180\,{\rm MHz}\right)^{-2.6}$~K, consistent with \cite{Dewd}.

Inspired by design plans for the Square Kilometer Array, we first consider 
a future ground-based interferometer with access to the final stages of the 
dark ages, $z\sim20-30$, with a baseline of $D= 6$ km [corresponding 
to a maximum angular scale $\ell_{\rm cover}(\nu)\sim 5800$ at redshift $z=30$], 
with $f_{\rm cover}=0.02$, surveying a sky fraction $f_{\rm sky}=0.75$ 
for a total of five whole years. As for the bandwidth, we surveyed a range between $\Delta\nu=0.1$ MHz and 10 MHz and found that $\Delta\nu\sim1$ MHz is the optimum value (for smaller bandwidths the noise $C_\ell$s dominate over the signal and for larger ones the number of redshift slices is too small).

For more optimistic constraints, we set $D= 50$ km, 
$f_{\rm cover}=0.1$, and assume ten whole years of observations. 
In order to get a result closer to the cosmic-variance limit we could perform the analysis from $z=20$, going up to the beginning of the dark ages, $z=200$. However, we find that it does not improve the results significantly, due to the rise of synchrotron radiation at low frequencies, which grows much more rapidly than the signal. We consider then the same redshift range as before, $z$ from 20 to 30.

One of the great advantages of 21-cm as a probe is the ability to analyze the tomography of the signal, enabling us to coadd information from different redshift slices. 
Summing over redshift slices, the signal-to-noise ratio is given by
\be
\left( S/N\right) =  \left[\sum_z \frac{f_{\rm sky}}{2}
     \sum\limits_{\ell_{\rm min}}^{\ell_{\rm max}}
     \frac{(2\ell+1)\left({C}^{\Vdm}_\ell(z)\right)^2}{\left(
     C^{\rm std}_{\ell}(z)+C^N_{\ell}(z)\right)^2}\right]^{\frac{1}{2}},
     \label{eq:snz}
\ee

\begin{figure}[htbp!]
	\includegraphics[width=1.\linewidth]{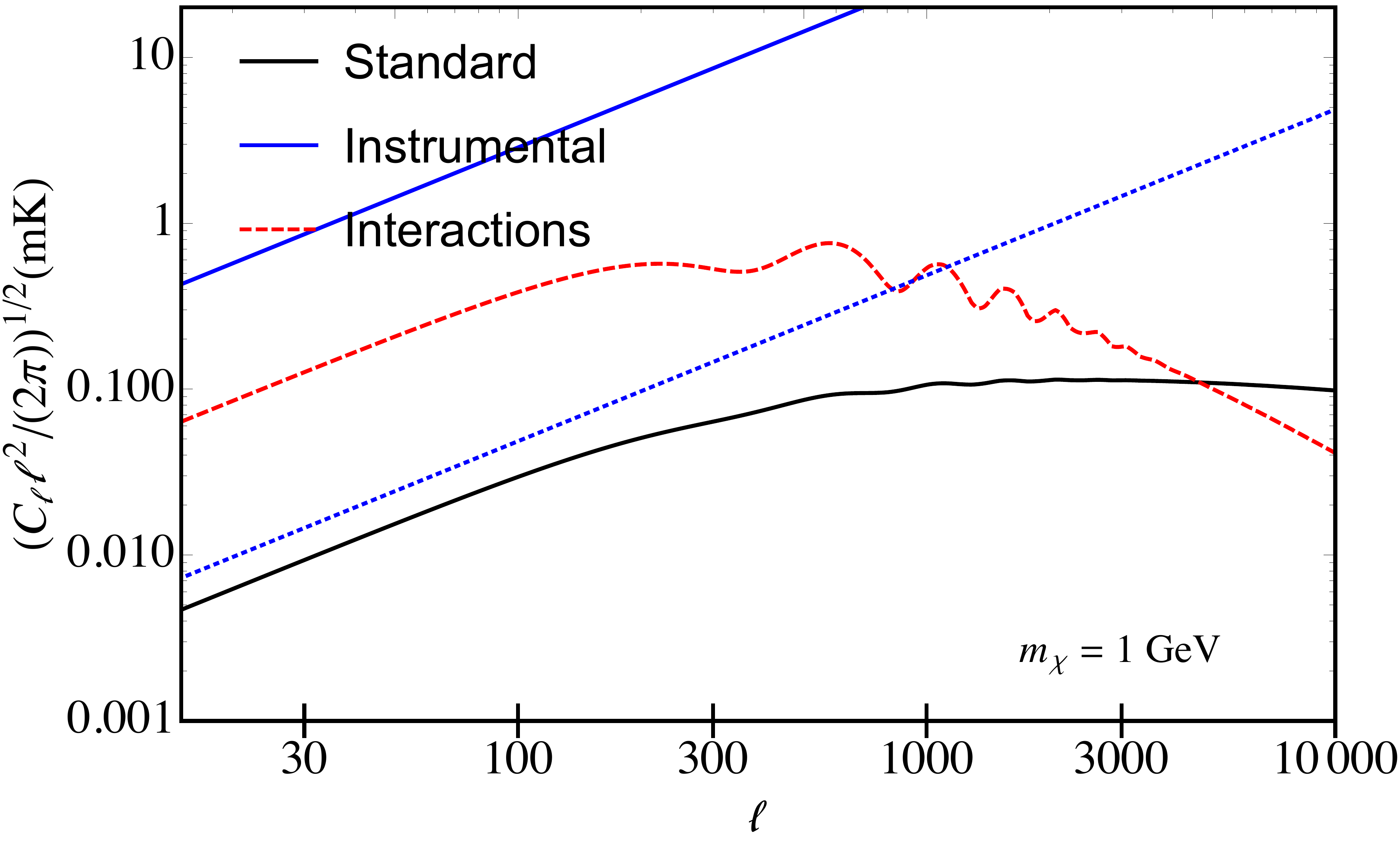}
	\caption{Angular power spectra at redshift $z=30$ with bandwidth $\Delta\nu=1$ MHz. In solid-black curve we show the usual primordial perturbations, in solid- and dotted-blue curves the instrumental noises for the realistic and optimistic cases [see Eq.~(\ref{eq:NoisePS}) and discussion below] and in dashed-red curve the new piece due to interactions for $\sigma_{41}=1$ and $\mdm= 1$ GeV.}
	\label{fig:Cldmb}
\end{figure}

In Fig.~\ref{fig:Cldmb} we show the $C_\ell$s for the usual primordial perturbations ($C_\ell^{\rm std}$), for the instrumental noise ($C_\ell^N$, both with next-generation and futuristic parameters), and for the new contribution due to interactions ($C_\ell^{\Vdm}$), all of them at redshift $z=30$.

\subsubsection*{Results}

Let us start by considering the realistic noise case (that corresponds to the experimental parameters of SKA) and find what the signal-to-noise ratio would be for detecting $\sigma_{41}=1$.
We calculate the signal-to-noise ratio for $\sigma_{41}=1$ in each redshift bin between $z=20$ and $z=30$ with Eq.~(\ref{eq:snz}).
We find the total signal-to-noise ratio to be S/N$\sim 3$ for the case of $ \mdm= 0.1$ GeV, S/N$\sim 9$ for  $\mdm=1$ GeV, and S/N$\sim 0.2$ for  $\mdm=10$ GeV. We could alternatively express the results in terms of the smallest $\sigma_{41}$ that would still give us a signal-to-noise ratio of 1, taken to be approximately $\sigma_{41,\rm min} = 1/\sqrt{S/N}$. We show the minimum detectable cross sections in Tab.~\ref{tab:min}.

Let us now move on to trying to find the smallest possible $\sigma_{41}$ detectable at $S/N=1$ in the more optimistic case.
In principle the amplitude $A$ of $C_{\ell}^{\Vdm}$, equal to $\bar T_{21,\rm rms}^2$, is a non-trivial function of redshift and $\sigma_{41}$. However, we find that for small values of $\sigma_{41}$ ($\sigma_{41}\lesssim 0.1$), the quantity $f(z)\equiv \bar T_{21,\rm rms}/\sigma_{41}$ is approximately independent of $\sigma_{41}$ (although it does depend on $\mdm$). Then we can construct an estimator for $\sigma_{41}$ for each redshift slice,
\be
\left(\widehat{\sigma^2_{41}}\right)_z = \dfrac{(\hat A)_z}{f^2(z)},
\ee
with variance given by $\sigma^2_{(\sigma^2_{41})_z} = \sigma^2_A(z)/f^4(z)$. We can then combine all the estimators into a minimum-variance one, finding the variance of the final redshift-independent estimator,
\be
\dfrac 1 {\sigma^2_{\sigma^2_{41}}}  = \sum_z \dfrac{f^4(z)}{\sigma^2_A(z)}.
\ee

With the optimistic experimental parameters defined above we find that the minimum $\sigma_{41}$ observable at 68\% C.L. (1$\sigma$) is $\sigma_{41}\lesssim 1.7 \times 10^{-3}$ for $\mdm=0.1$ GeV, $\sigma_{41}\lesssim  4.3 \times 10^{-3}$ for $\mdm=1$ GeV, and $\sigma_{41}\lesssim  3.6 \times 10^{-2}$ for $\mdm=10$ GeV. These results are about 2 orders of magnitude better than the CMB constraints found in \cite{1311.2937}, where $\sigma_{41}\lesssim  16 (\mdm/10\, \rm GeV)$.

\begin{center}
\begin{table}[h]
    \begin{tabular}{| l | c | c | c | }
    \hline
    $\mdm$ [GeV]&  1/10 &  1   & 10  \\             
   \hline
     Fluctuations (realistic) & 6$\times 10^{-42}$ & 3$\times 10^{-42}$ & 2$\times 10^{-41}$  \\         
     Fluctuations (optimistic)  & 2$\times 10^{-44}$ & $4 \times 10^{-44}$ & $4\times 10^{-43}$ \\  
     Global signal (1\% error) & 4$\times 10^{-43}$ & 1$\times10^{-42}$ & $2\times 10^{-42}$ \\            
	\hline
    \end{tabular}
\caption{Minimum $\sigma_{0}$ (in cm$^2$, corresponding to $\sigma_{41} \times 10^{41}$)  detectable with both realistic and optimistic interferometer parameters at 68\% C.L., as well as with global-signal analysis with $1\%$ accuracy for three different dark-matter masses $\mdm$ (in GeV).}
\label{tab:min}
\end{table}
\end{center}

\section{Discussion}
\label{sec:discussion}

Before concluding we would like to make a few remarks:

$\bullet$ As we have shown, interactions between dark matter and baryons give rise to a new heating term, which can increase the temperature of the baryons significantly. We only used that heating to study dark-ages physics but this result may have applications beyond our analysis, for example in the epoch of reionization \cite{1508.04138,1402.0940}.

$\bullet$ In this work we have focused only on the case where $\sigma\sim v^{n}$ with $n=-4$, but one may wonder whether the dark ages can potentially provide new information not contained in the CMB analysis for other values of $n$. 
Since the dark ages occur more recently than decoupling, we have only been interested in interactions that increase at later times. Ref.~\cite{1311.2937} showed that the interaction rate grows for $n\leq -3$,
so all results that we could forecast for $n>-3$ would be worse than those obtained with CMB studies. That still leaves $n=-3$ as a potential interaction to study, for example.

$\bullet$ It is also worth mentioning that if we wanted to translate these results to a constraint specific to a dark-matter milicharge model \cite{astro-ph/0406355}, the ionization fraction of the baryons would cause a suppression of $x_e\sim 10^{-4}$.

\begin{figure}[htbp!]
\includegraphics[width=1.\linewidth]{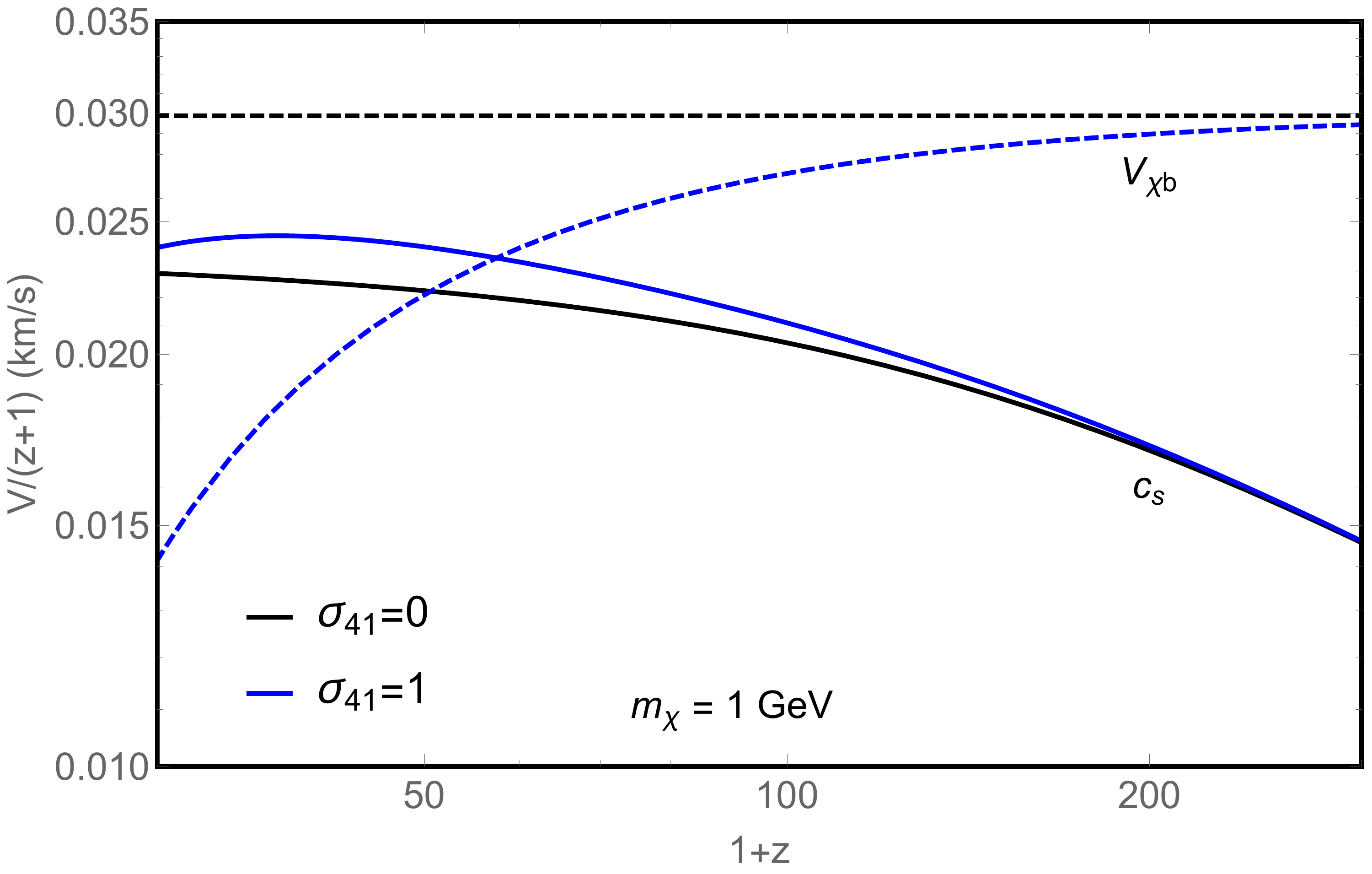}
\caption{Values of the relative velocity (solid curves) and the thermal speed of sound (dashed curves) divided by $(1+z)$. We show the collisionless case (black curves) and the case with $\sigma_{41}=1$ (blue curves), for $\mdm=1$ GeV.} 
\label{fig:v}
\end{figure}

$\bullet$  We have also found a decrease in the bulk relative velocity of baryons and dark matter characterized by a drag, Eq.~(\ref{eq:drag}). In Fig.~\ref{fig:v} we show the unperturbed relative velocity $\Vdm$, found by solving Eqs.~(\ref{eq:Tchi})-(\ref{eq:xe}) with initial relative velocity $V_{\chi b,0}=V_{\rm rms}$, and baryon speed of sound $c_s\equiv \sqrt{3T_b/m_b}$. We also plot the same two velocities for an interacting case. All velocities are divided by a factor of $1/(z+1)$ to eliminate a fiducial redshift dependence.

In the standard case the speed of sound is always below the bulk one, which creates supersonic flow of the baryons \cite{1005.2416,1312.4948}. Including collisions can both raise the thermal velocity as well as decrease the relative one, so it reduces the Mach number $\cal N$ to be lower than 1 at lower redshifts, which could affect the formation of small-scale structure \cite{1201.3614}.

\begin{figure}[htbp!]
\includegraphics[width=1.1\linewidth]{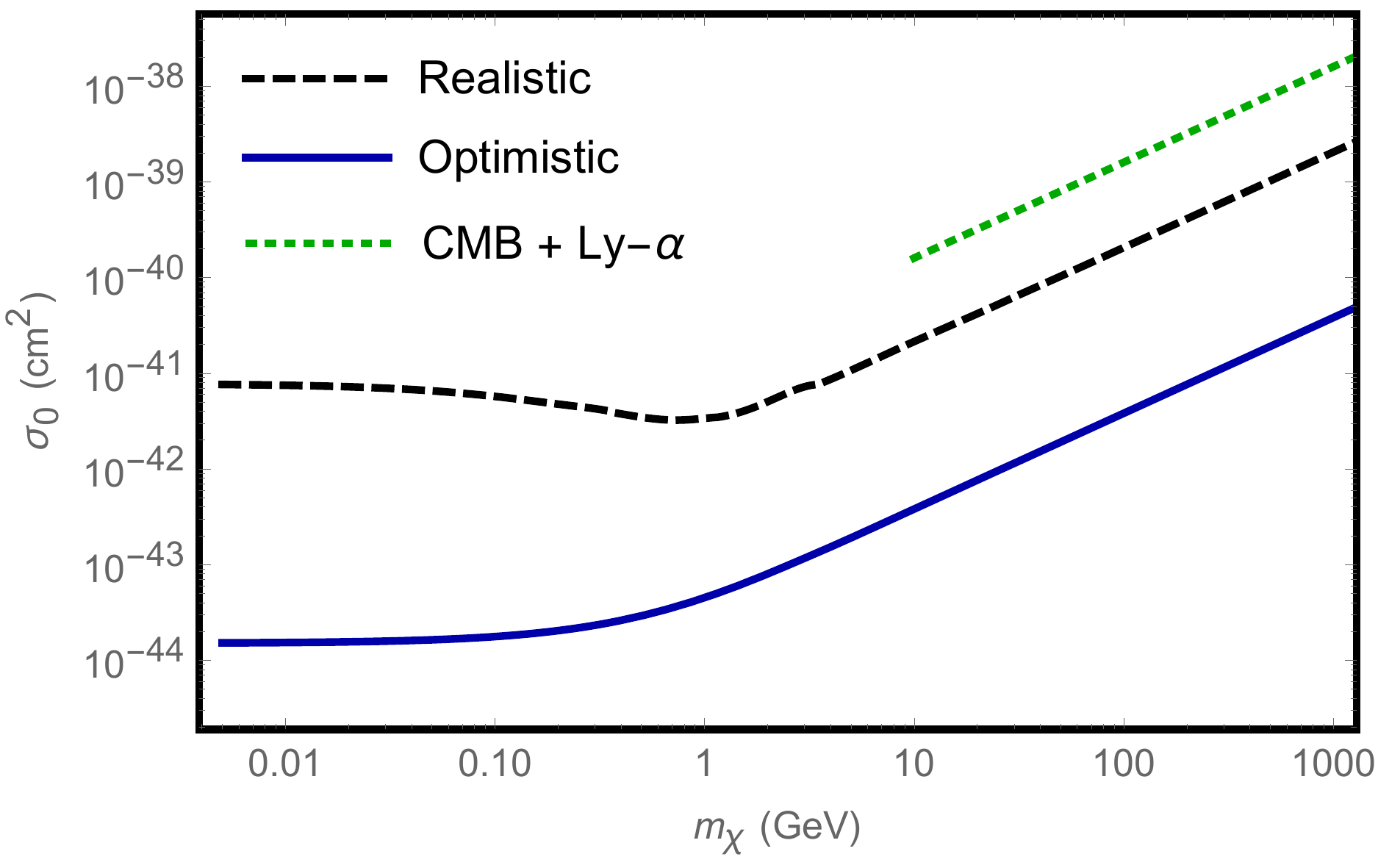}
\caption{Minimum $\sigma_{0}$ (in cm$^2$) detectable as a function of $\mdm$ in GeV. In black curve we show the results for the case with realistic parameters and in blue curve the one with optimistic parameters. In dotted-green curve we display the current CMB constraint (only valid for $\mdm\gg$ GeV).} 
\label{fig:sigma0}
\end{figure}

$\bullet$ Finally, throughout the text we have quoted results for $\mdm=0.1$, 1, and 10 GeV. For lower masses, the result is independent of mass, and for higher masses it depends on $\sigma_0/\mdm$. We show a larger range of dark-matter masses in Fig.~\ref{fig:sigma0}, where we plot the minimum $\sigma_{0}$ one could detect at a signal-to-noise ratio of 1, as a function of the dark-matter mass $\mdm$. We show how the result asymptotes for very high and very low $\mdm$, and we also compare with the CMB+Ly$\alpha$ analysis in Ref.~\cite{1311.2937}, shown in dotted-green curve, which is only valid for large $\mdm$.

\section{Conclusions}
\label{sec:conc}

We have shown that adding interactions between dark-matter and baryons (in a velocity-dependent way $\sigma=\sigma_0 v^{-4}$) can dramatically change  the behavior of the baryons during the dark ages. Intuitively it would seem that coupling the two fluids will tend to decrease the temperature of the baryons in favor of the dark-matter temperature. We have proven, however, that there is an extra heating term that appears due to the relative velocity between the two fluids, which tends to convert initial kinetic energy into thermal energy. 
For a wide range of non-zero initial relative velocities and dark-matter masses $\mdm\gtrsim 1$ GeV, we find that the heating dominates over the cooling of the baryons (if the initial velocity is zero, we find that there is only cooling, as expected).

Heating up the baryons affects the physics of the dark ages significantly. A higher baryon temperature will translate into a higher spin temperature during the dark ages, and hence a smaller brightness temperature of the 21-cm line (if one cools down the baryons, the effect is the opposite). The dependence of the heating on the initial relative velocity makes the brightness temperature depend on the position in the sky, and hence
creates an extra source of perturbations to the brightness temperature, sourced by the relative velocity perturbations. We have calculated the power spectrum of these perturbations and compared it to the standard primordial one.

To find constraints to dark-matter--baryon interactions we have studied two probes. 
First, the global signal during the dark ages gets modified by the interactions, and assuming an experiment that could detect the global signal at peak redshift with an accuracy of 1\% we can forecast $\sigma_{0}\lesssim 10^{-42}$ cm$^2$.
Second, the interactions create a new contribution to the power spectrum. We studied the case of a realistic ground-based interferometer focusing on the end of the dark ages ($z= 20-30$) and found that a cross section of  $\sigma_{0}\sim 10^{-41}$ cm$^2$ could be marginally detected. We also considered a more futuristic experiment and found that the minimum cross section one could measure is  $\sigma_{0}\sim 10^{-44}$ cm$^2$, more than 2 orders of magnitude better than can be achieved by CMB+Ly$\alpha$ analysis, and with a broader mass range.

\begin{acknowledgments}
	It is our pleasure to thank Joe Silk, Anastasia Fialkov, Shawn Westerdale, and especially Marc Kamionkowski for useful discussions. 
	This work was supported by NSF Grant No. 0244990, NASA NNX15AB18G, the John Templeton Foundation, and the Simons Foundation.
\end{acknowledgments}

\newpage

\appendix 
\onecolumngrid

\section{Heating rate}
\label{App:heat}

The baryon heating per unit time is given by
\ba
\dfrac{dQ_b}{dt} =& \dfrac{m_b \rho_\chi}{(\mdm+m_b)} \int d^3 v_b f_b \int d^3 \vdm f_\chi (\vdm) \bar \sigma\left(|\mathbf \vdm - \mathbf v_b|\right)  |\mathbf \vdm - \mathbf v_b| \left[ \mathbf v_{\rm CM}\cdot  (\mathbf v_b-\mathbf v_\chi)\right],
\end{align}
we again resort to writing everything in terms of $v_m$ and $v_-$, this time defining them to be thermal, so they do not depend on $\Vdm$, and we use the general expression $\bar\sigma=\sigma_0 v^n$, so that
\be
\dfrac{dQ_b}{dt} =\dfrac{ m_b \rho_\chi\sigma_0}{(\mdm+m_b)} \int d^3 v_m f_m \int d^3 v_- f_- \left(|\mathbf\Vdm + \mathbf v_-|\right)^{n+1}  \left[ \mathbf v_{\rm CM}\cdot(\mathbf\Vdm + \mathbf v_-)\right],
\ee
and we can calculate the scalar product term using $\mathbf v_{\rm CM} =a \mathbf \Vdm  +b\mathbf v_-+\mathbf v_m $, with $a= \mdm/(\mdm+m_b)$ and  $b=(T_\chi-T_b)/[u_{\rm th}^2 (\mdm+m_b)]$.
Then the scalar product will be $\left( \mathbf v_{\rm CM}\cdot \mathbf v_-\right) = a \mathbf\Vdm^2 + b\mathbf v_-^2 + (a+b) \mathbf \Vdm \cdot \mathbf v_- + \mathbf v_m \cdot (\dots)$.
We will have to integrate over $v_m$ and $v_-$, which makes it obvious that the factors proportional to $v_m$ will cancel out, as
\begin{align}
&\int d^3 v_m f_m = 1 \\
&\int d^3 v_m f_m  \mathbf v_m \cdot  \mathbf A = 0.
\end{align}
Therefore
\be
\dfrac{dQ_b}{dt} =  \dfrac{m_b \rho_\chi \sigma_0}{(\mdm+m_b)} \int d^3 v_- f_- \left(|\mathbf\Vdm + \mathbf v_-|\right)^{n+1}  \left( a \mathbf \Vdm^2 + (a+b)\mathbf v_- \cdot \mathbf \Vdm + b \mathbf v_-^2 \right),
\ee
so we can calculate the heating rate in terms of two integrals,
\be
\dfrac{dQ_b}{dt} = \dfrac{m_b \rho_\chi \sigma_0}{(\mdm+m_b)} \left[ a I_1(n)+ b I_2(n) \right],
\ee
defining
\begin{align}
I_1(n)&=\dfrac{\Vdm^{n+6}}{(2\pi)^{1/2}u_{\rm th}^3} \int_0^\infty dx x^2 e^{-x^2 r^2/2 } \int_{-1}^1 dy \left( 1+x^2 + 2 x y \right)^{(n+1)/2} (1+x y),\quad \rm{ and }
\nonumber \\
I_2(n)&=\dfrac{\Vdm^{n+6}}{(2\pi)^{1/2}u_{\rm th}^3} \int_0^\infty dx x^2 e^{-x^2 r^2/2 } \int_{-1}^1 dy \left( 1+x^2 + 2 x y \right)^{(n+1)/2} x (x+y),
\end{align}
with $x=v_-/\Vdm$ and $r=\Vdm/u_{\rm th}$. We can perform the two different $y$ integrals to find,
\begin{align}
&\int_{-1}^1 dy \left( 1+x^2 + 2 x y \right)^{(n+1)/2} (1+xy) =\frac{(x-1) (n+x+4) \left| x-1\right| ^{n+3}+(n-x+4) (x+1)^{n+4}}{(n+3) (n+5) x}, \quad \rm{ and }
\nonumber
 \\
&\int_{-1}^1 dy \left( 1+x^2 + 2 x y \right)^{(n+1)/2} (x+y) =\frac{(x+1)^{n+4} [(n+4) x-1]-(x-1)^3 [(n+4) x+1] \left| x-1\right| ^{n+1}}{(n+3) (n+5) x^2},
\end{align}
which means that
\begin{align}
I_1(n)&=\dfrac{\Vdm^{n+6}}{(2\pi)^{1/2}u_{\rm th}^3} \int_{-\infty}^\infty dx  e^{-x^2 r^2/2 } x \frac{(x-1) (n+x+4) \left| x-1\right| ^{n+3}}{(n+3) (n+5)}, \quad \rm{ and }
\nonumber \\
I_2(n)&=-\dfrac{\Vdm^{n+6}}{(2\pi)^{1/2}u_{\rm th}^3} \int_{-\infty}^\infty dx e^{-x^2 r^2/2 } x \dfrac{(x-1)^3 [(n+4) x+1] \left| x-1\right| ^{n+1}}{(n+3) (n+5)}.
\end{align}
These functions can be expressed in terms of hypergeometric functions, but since we are interested in the $n=-4$ case, let us solve just for that
\begin{align}
I_1(-4)&=-\dfrac{\Vdm^{2}}{(2\pi)^{1/2}u_{\rm th}^3} \int_{-\infty}^\infty dx  e^{-x^2 r^2/2 } x^2 \frac{\left| x-1\right|}{(x-1)} = -\dfrac{1}{(2\pi )^{1/2}u_{\rm th}} \frac{2 e^{-\frac{r^2}{2}} r-\sqrt{2 \pi } \text{erf}\left(\frac{r}{\sqrt{2}}\right)}{r}, \quad \rm{ and }
\nonumber\\
I_2(-4)&=\dfrac{\Vdm^{2}}{(2\pi)^{1/2}u_{\rm th}^3} \int_{-\infty}^\infty dx e^{-x^2 r^2/2 } x \dfrac{\left| x-1\right|}{(x-1)} = \dfrac{1}{(2\pi )^{1/2}u_{\rm th}} 2 e^{-\frac{r^2}{2}}.
\end{align}

We finally find,
\be
\dfrac{dQ_b}{dt} = \dfrac{m_b \rho_\chi \sigma_0}{(\mdm+m_b)} \dfrac{1}{(2\pi )^{1/2}u_{\rm th}} \left[2 (b-a)e^{-\frac{r^2}{2}} + a \dfrac{\sqrt{2 \pi}}{r} \text{erf}\left(\frac{r}{\sqrt{2}}\right) \right],
\ee
or, plugging $b$ and $a$,
\ba
\dfrac{dQ_b}{dt} &= \dfrac{m_b \rho_\chi \sigma_0}{(\mdm+m_b)^2\sqrt{2\pi} u_{\rm th}} \left[2 \dfrac{T_\chi - T_b}{u_{\rm th}^2}e^{-\frac{r^2}{2}} - \mdm \dfrac{F(r)}{r} \right],
\label{eq:solution}
\end{align}
which in the $v_-\gg \Vdm$ limit (corresponding to $r\to0$), yields
\be
\dfrac{dQ_b}{dt} \approx 2b \dfrac{m_b \rho_\chi \sigma_0}{(\mdm+m_b)} \dfrac{1}{(2\pi )^{1/2}u_{\rm th}} =  2\dfrac{m_b \rho_\chi \sigma_0}{(\mdm+m_b)^2} \dfrac{1}{(2\pi )^{1/2}}  {\left( \frac{T_\chi}{\mdm} + \frac{T_b}{m_b} \right)^{-3/2}} \left (T_\chi - T_b \right).
\ee
This expression does not contain a temperature-independent heating (as expected), and matches exactly Ref.~\cite{1311.2937} for $n=-4$. However, Eq.~(\ref{eq:solution}) works for all $r$. In the opposite limit, where $\Vdm\gg v_-$ (and then $r\to\infty$), we find that the heating is
\be
\dfrac{dQ_b}{dt} \approx  \dfrac{\mdm m_b \rho_\chi \sigma_0}{(\mdm+m_b)^2 \Vdm},
\ee
which indeed does not cancel even for $T_\chi=T_b$, and ends up being independent of temperature.

\end{document}